\documentclass[acmsmall, authorversion,nonacm]{acmart}
\AtBeginDocument{%
	}

\usepackage{makecell}
\usepackage{tikz}
\usepackage{amsmath}
\usepackage{booktabs}
\usepackage{filecontents}
\usepackage{xspace}
\usepackage{multirow} 
\usepackage{graphicx}
\usepackage{caption}
\usepackage{forest}
\usepackage{subcaption}
\usepackage{wrapfig}
\usepackage{xcolor,colortbl}
\usepackage{etoolbox} 
\usepackage[compact]{titlesec}
\usepackage{cleveref}

\crefformat{section}{\S#2\color{blue}#1#3} 
\crefformat{subsection}{\S#2\color{blue}#1#3}
\crefformat{subsubsection}{\S#2#1#3}

\renewcommand{\paragraph}[1]{\noindent\textbf{#1}}
\titlespacing{\paragraph}{0pt}{*0.5}{*1}
\usepackage{enumitem}
\setlist[enumerate]{noitemsep, topsep=0pt}
\setlist[itemize]{noitemsep, topsep=0pt}
\titlespacing\section{0pt}{*1.8}{*1.1}
\titlespacing\subsection{0pt}{*1.5}{*1.1}
\titlespacing\subsection{0pt}{*1.3}{*0.8}
\usepackage{subcaption}
\usepackage{caption}
\begin{document}
	\newcommand{\chase}[1]{\textcolor{blue}{[#1 - Chase]}}
	\newcommand{\prs}[1]{\textcolor{green}{[#1 - PRS]}}
	\newcommand{\aaron}[1]{\textcolor{orange}{[#1 - AGJ]}}
	\newcommand{\ab}[1]{\textcolor{purple}{[#1 - AB]}}
	\newcommand{\eg}{{\it e.g.}}
	\newcommand{\ie}{{\it i.e.}}
	\title{NetDiffusion: Network Data Augmentation Through Protocol-Constrained Traffic Generation}
	\author{Xi Jiang}
	\affiliation{%
		\institution{The University of Chicago}
		\city{Chicago}
		\country{USA}}
	\email{xijiang9@uchicago.edu}
	
		\author{Shinan Liu}
	\affiliation{%
		\institution{The University of Chicago}
		\city{Chicago}
		\country{USA}}
	\email{shinanliu@uchicago.edu}
	
		\author{Aaron Gember-Jacobson}
\affiliation{%
	\institution{Colgate University}
	\city{Hamilton}
	\country{USA}}
\email{agemberjacobson@colgate.edu}
		\author{Arjun Nitin Bhagoji}
\affiliation{%
	\institution{The University of Chicago}
	\city{Chicago}
	\country{USA}}
\email{abhagoji@uchicago.edu}
		\author{Paul Schmitt}
\affiliation{%
	\institution{University of Hawaii, Manoa / Invisv}
	\city{Hawaii}
	\country{USA}}
\email{pschmitt@hawaii.edu}
		\author{Francesco Bronzino}
\affiliation{%
	\institution{École Normale Supérieure de Lyon}
	\city{Lyon}
	\country{France}}
\email{francesco.bronzino@ens-lyon.fr}
		\author{Nick Feamster}
\affiliation{%
	\institution{The University of Chicago}
	\city{Chicago}
	\country{USA}}
\email{feamster@uchicago.edu}

	\begin{abstract}
	Datasets of labeled network traces are essential for a multitude of machine
	learning (ML) tasks in networking, yet their availability is hindered by
	privacy and maintenance concerns, such as data staleness. To
	overcome this limitation, synthetic network traces can often augment
	existing datasets. Unfortunately, current synthetic trace generation methods,
	which typically produce only aggregated flow statistics or a few selected packet
	attributes, do not always suffice, especially when model training relies
    on having features that are only available from packet traces. This shortfall
	manifests in both insufficient statistical resemblance to real traces and
	suboptimal performance on ML tasks when employed for data augmentation.
	In this paper, we apply
	diffusion models to generate high-resolution 
	synthetic network traffic traces. We present \emph{NetDiffusion},
	a tool that uses a finely-tuned, controlled variant of a Stable Diffusion
	model to generate synthetic network traffic that is high fidelity and
    conforms to protocol specifications.
	Our evaluation demonstrates that 
	packet captures generated from NetDiffusion can achieve higher statistical similarity to real
	data and improved ML model performance than current
    state-of-the-art approaches (e.g., GAN-based approaches). Furthermore,
	our synthetic traces are compatible with
	common network analysis tools
	and support a myriad of network tasks,
	suggesting that NetDiffusion can serve a broader spectrum of network analysis and testing tasks, extending beyond ML-centric applications.
\end{abstract}

	\maketitle
	\section{Introduction}\label{sec:intro}

Modern networks are increasingly reliant on machine learning (ML) techniques for a wide range of management tasks, ranging from security to performance optimization.
A central impediment when training network-focused ML models is the scarcity of labeled network datasets,
as their collection and sharing are often associated with high costs and privacy concerns,
particularly when data is collected from real-world networks~\cite{sommer2010outside, mahoney2003network,mchugh2000testing,abt2014we,de2023survey}.
Unfortunately, existing public datasets rarely receive updates,
making them static and unable to reflect evolving network behaviors~\cite{kenyon2020public,ring2019survey,labovitz2010internet}.
These limitations hinder the ability to train robust ML models that accurately reflect evolving real-world network conditions.

These challenges can be addressed through the creation of new synthetic network traces based on existing datasets.
This approach aims to preserve the inherent characteristics of network traffic while introducing variations,
thereby enhancing dataset size and diversity~\cite{sharafaldin2018toward,xu2021stan,nukavarapu2022miragenet,yin2022practical,sommers2004harpoon,vishwanath2009swing,lin2020using,zander2005kute}.
Unfortunately, current state-of-the-art synthetic trace generation methods,
particularly those based on Generative Adversarial Networks (GANs)-based
methods~\cite{ring2019flow,yin2022practical,lin2020using,xu2021stan,xu2019modeling},
are not always sufficient for producing high-quality synthetic network traffic.
Specifically, these approaches tend to focus on a limited set of attributes or statistics,
as early machine learning for network tasks often relied on basic flow
statistics for classification~\cite{paxson1994empirically,dewes2003analysis,claffy1995internet,
	lang2003synthetic,lang2004synthetic,bernaille2006early,karagiannis2005blinc}.
With recent ML advancements utilizing detailed raw
network traffic to achieve enhanced classification accuracy~\cite{rimmer2017automated,zheng2022mtt,
	lotfollahi2020deep,akbari2021look,yao2019identification,wang2020encrypted,
	shapira2019flowpic,cui2019session,bu2020encrypted,ma2021encrypted,sun2020encrypted}, there is a clear need for synthetic traffic generation
that includes the intricate, potentially unforeseen patterns present in full network traces. Existing traffic generation methods face two main issues:
(1)~a lack of statistical similarity with real data
due to the limited attributes in existing methods,
making the synthetic data highly sensitive to variations,
and (2)~unsatisfactory classification accuracy when synthetic statistical
attributes are used to augment existing datasets.
Moreover, their simplistic attribute focus and disregard for transport and network layer protocol behaviors prevent their use with traditional networking tools such as tcpreplay~\cite{tcpreplay2023} or Wireshark~\cite{beale2006wireshark}.

Fortunately, the general increase in available computational power and the breakthroughs in 
high-resolution image generation techniques,
particularly diffusion models~\cite{sohl2015deep, rombach2022high, ramesh2022hierarchical}, 
offer a promising avenue to overcome these challenges.
In contrast to GANs, diffusion models are able to capture both broad patterns and detailed dependencies.
This inherent generative quality makes them an ideal choice 
for producing network traces with high statistical resemblance
to real traffic and full packet header values. By incorporating 
conditioning techniques, diffusion models can generate structured data
that conforms to specific network properties, which ensures the desired 
sequential inter-packet characteristics and rough protocol dependencies.
Moreover, the gradient dynamics of the training process in diffusion 
models is much more stable than GANs. 
We discuss detailed benefits of diffusion models in Section~\ref{sec:method}.
These attributes collectively position diffusion models as a compelling choice for advancing the
state-of-the-art for synthetic network trace generation, addressing the extant limitations of current methodologies.

In this paper, we introduce NetDiffusion, an approach to synthetic
raw network traffic generation for producing packet headers
leveraging fine-tuned,
controlled stable diffusion models.
Our contributions are as follows:

\begin{enumerate}
	\item \textbf{Generation of synthetic network traces
		with high resemblance to real traffic:}
	      Using stable-diffusion techniques, we propose a two-fold strategy:
	      (1) a conversion process for transforming raw packet captures to image representations (and vice versa),
	      and (2) fine-tuning a text-to-image diffusion model based on packet capture-converted
	      images for generating synthetic packet captures.
	      To improve resemblance to real network traffic,
	      we employ controlled generation techniques to maintain
	      fidelity to the protocol and header field value distributions
	      observed in real data and, post generation,
	      use domain knowledge-based heuristics
	      to finely check and adjust the generated fields,
	      ensuring their semantic correctness in terms of
	      compliance with transport and network layer protocol rules.

	\item \textbf{Improved classification accuracy in ML scenarios with
	synthetic network traffic augmentation:}
	      We conduct a case study evaluation on a curated traffic classification dataset.
	      By integrating NetDiffusion-generated network traffic
	      into the real dataset at varying proportions during training and testing,
	      we observe a general increase in accuracy compared to the state-of-the-art
	      generation method~\cite{yin2022practical}.
	      This improvement is attributed to our synthetic data's significantly
	      high statistical resemblance
	      to the real dataset.
	      Additionally, our method shows promise in addressing class imbalance issues,
	      enhancing the accuracy of ML models in such cases.
	   
	\item \textbf{Extended applicability of synthetic network traffic for network analysis and testing
		      beyond ML tasks:}
		      NetDiffusion-generated network traffic can be converted into raw packet captures
		      suitable for traditional network analysis and testing tasks.
		      We validate this compatibility through tests with tools
		      such as Wireshark and Scapy~\cite{rohith2018scapy},
		      as well as tcpreplay for retransmission.
		      More importantly, we show that critical statistical features
		      for various network operations can be effectively extracted
		      from the generated network traffic.

\end{enumerate}

	\section{Motivation}\label{sec:motivation}
The use of publicly available network datasets has significantly aided advancements in
applying ML to networks,
as well as network analysis and testing methodologies.
For example, models trained on network datasets have been helpful in tackling challenges like anomaly detection,
traffic classification, and network optimization, which in turn enhances network security and performance ~\cite{bronzino2021traffic,jiang2021automating,liu2023grounding,lotfollahi2020deep,jiang2023ac,jiangtowards,abt2014we,moustafa2019holistic,kavitha2021network,thomas2018survey,cholakoska2021analysis}.
Additionally, these datasets are valuable for network analysts, aiding in understanding network behaviors,
identifying performance issues,
and evaluating the performance of network security tools like firewalls and intrusion detection systems~\cite{vasudevan2011ssenet,negandhi2019intrusion,rathore2016real}.

\subsection{Network Data Scarcity}
Well-known network datasets such as CAIDA~\cite{caida2019}, MAWI~\cite{cho2000traffic}, UNSW-NB15~\cite{7348942},
and KDD~\cite{kddcup1999,tavallaee2009detailed} have been essential for numerous research projects in network science.
However, the lack of updated datasets often hinders further progress.
Those with the means to capture large-scale traffic,
typically network operators and organizations
with specialized hardware and network,
are often hesitant to share their data due to the risk
of exposing sensitive or personally identifiable information (PII).
Even when entities are amenable to sharing,
the task of providing consistent updates and ensuring reliable labels for sanitized data is daunting.
Labeling network data is inherently challenging because of its dynamic nature,
such the continuous evolution of network behaviors and threats.
Notably, the CAIDA, UNSW-NB15, KDD Cup 99~\cite{kddcup1999}, and
NSL-KDD~\cite{tavallaee2009detailed} datasets
were last updated in 2020, 2015, 1999, and 2009 respectively,
revealing notable gaps in data recency
which render them less reflective of evolving network dynamics.
Even frequently updated datasets like MAWI~\cite{cho2000traffic} are not
exempt from issues, with instances of missing data from hardware failures
and substantial duplicate traces.
While not an exhaustive list of datasets, the issues highlighted are common across the
board, accentuating the need for newer data to fuel ongoing network research and analysis.

\subsection{Data Augmentation Using Synthetic Data}
Data augmentation through synthetic data has proven effective in many fields.
In computer vision, for instance, synthetic images have improved model performance,
especially when there's a shortage of labeled data~\cite{meister2021synthetic,chlap2021review,talukdar2018data,shin2018medical,liu2021synthetic}.
The success of synthetic data augmentation is largely
attributed to generative methods which have
showcased remarkable versatility in a variety of domains:
In medical imaging, GANs
have been harnessed to augment datasets, significantly
enhancing the performance of diagnostic models~\cite{frid2018gan,guan2022medical,chen2022generative}.
In the domain of natural language processing, Variational
Autoencoders (VAEs) have been utilized to create synthetic text
data, aiding in tasks such as sentiment analysis and language
translation~\cite{wang2019vector,tian2020learning,chorowski2019unsupervised}. In audio processing, the advent of WaveGAN has facilitated
the expansion of sound datasets, proving indispensable for
applications like speech recognition and sound event detection~\cite{donahue2018adversarial,donahue2018synthesizing,barahona2020synthesising}.

Translating these successes to the networking domain,
certain endeavors have emerged, attempting to augment
network datasets through the generation of synthetic network data~\cite{ring2019flow,yin2022practical,lin2020using,xu2021stan,xu2019modeling}.
A notable state-of-the-art attempt in this regard
is NetShare~\cite{yin2022practical} which utilizes GANs to produce IPFIX~\cite{rfc7012}/NetFlow
\cite{claise2004cisco}-style
statistics on network traffic. For simplicity, we refer to this general type of
aggregated statistucal attributes as NetFlow for the rest of the paper.
Yet, its focus on
statistical properties might miss important network patterns essential for high
ML accuracy.
At the same time, the limited number of attributes that it focuses prohibits
the generation of comprehensive, raw network traffic such as packet captures, which are essential
for additional non-ML tasks such as network analysis and testing.
In this paper, we do not consider other
non-generative methods~\cite{lacage2006yet,henderson2008network,buhler2022generating,botta2012tool}
like TRex~\cite{ciscotrex2023} and NS-3~\cite{henderson2008network} because,
while useful for specific tasks, they lack the flexibility needed for broader dataset augmentation.
They often rely on predefined templates or rules,
which may not capture the evolving nature of network traffic or the complex interactions between
various network protocols and applications.

\subsection{Inadequate Performance from Existing Methods} \label{subsec:inadequate_existing}
We provide a short case study on NetShare,
which is the current state-of-the-art
network data generation method that
produces NetFlow attributes, \ie, derived
statistics from raw network traffic flows.
Following the method in the original paper,
we test the accuracy of a variety of
ML model--Random Forest (RF), Decision Tree (DT), and
Support Vector Machine (SVM)--under three scenarios:
(1) training and testing on real NetFlow data,
(2) training on NetShare synthetic data and testing on real data,
and (3) vice versa.
In this paper, we focus on a traffic classification
task using a curated dataset, detailed in Section~\ref{sec:evaluation}.
The classification task is divided into \textit{micro} and \textit{macro} levels.
On the micro level, the goal is to classify network traffic flows
into specific applications,
encompassing 10 distinct classes such as YouTube and Amazon.
On the macro level, the aim is to classify network
traffic flows into broader service categories,
spanning 3 classes, like streaming and web browsing.

\subsubsection{Unsatisafactory ML Accuracy.}
(1)~NetShare Limitations:
The results from Table~\ref{tab:motivation_model_performance} showcases
a clear drop in accuracy when models are trained or tested
on synthetic NetFlow data generated by NetShare
compared to when only real NetFlow data is used,
irrespective of the classification level.
This points to a likely shortfall of NetShare in preserving critical
distinguishing feature values inherent in the real dataset,
which adversely affects the models' ability to
accurately classify network traffic.
(2)~{NetFlow vs. Raw Traffic:}
Following the NetShare evaluation, we compared the accuracy
achieved with real NetFlow data to that when models train
and test on raw network traffic in pcap format.
This comparison underscores the information
loss encountered when using NetFlow data
and the potential classification performance
gain from leveraging raw network traffic:
The highest
accuracy is achieved with raw network traffic,
with SVM arriving at near-perfect accuracy.
Conversely, a noticeable decrease in accuracy is
observed with real NetFlow data, 
suggesting that its limited feature set adversely
affects classification accuracy.

\begin{wraptable}[10]{r}{0.6\columnwidth}
	\centering
	\small
	\resizebox{0.6\columnwidth}{!}{
		\begin{tabular}{|r|r|r|r|}
			\hline
			\multirow{2}{*}{\textbf{Training/Testing}} & \multirow{2}{*}{\makecell[r]{\textbf{Data Format}                                               \\ \textbf{(Generation Method)}}}& \multicolumn{2}{c|}{\textbf{Highest Accuracy (Model)}} \\
			
			&                                                   & \textbf{Macro-level} & \textbf{Micro-level} \\
			\hline
			\multirow{2}{*}{Real/Real}                 & Network traffic (pcap) (N/A)                      & 1.00 (SVM)           & 0.978 (SVM)          \\
			\cline{2-4}
			& NetFlow (N/A)                                     & 0.86 (RF)            & 0.648 (DT)           \\
			\hline
			\multirow{1}{*}{Synthetic/Real}
			& NetFlow (NetShare)                                & 0.396 (RF)           & 0.140 (SVM)          \\
			\hline
			\multirow{1}{*}{Real/Synthetic}
			& NetFlow (NetShare)                                & 0.503(SVM)           & 0.102 (RF)           \\
			\hline
	\end{tabular}}
	\caption{Comparison of model accuracy using real, 
	synthetic NetFlow data, and raw network traffic. 
	Results highlight a decrease in accuracy with 
	NetShare's synthetic data and a boost when using 
	raw traffic. Only the top-performing model is displayed.}
	\label{tab:motivation_model_performance}
\end{wraptable}
These observations lead to two main insights: First,
the need for synthetic data generation methods to effectively
preserve critical distinguishing feature values to maintain classification
accuracy; Second, the advantage of using raw network traffic
over NetFlow data due to its richer information content.
The push towards generating raw network traffic
to retain the fine-grained details and statistical properties
of real network traffic, appears to be a crucial step to overcome
the limitations observed with NetShare generated data and
NetFlow data.
\subsubsection{Limited Applicability to Non-ML tasks.}
Besides the performance of synthetic network data in ML tasks,
another important metric to validate its usefulness
is its applicability to traditional network analysis and testing tasks,
such as packet-wise analysis and replay.
This is important because, unlike other forms of data such as
images where the quality of the data can be inferred relatively
easily by visual resemblance to real images, it is hard for network
experts to manually examine raw network traffic to verify its quality.
NetFlow data, encapsulating aggregated or derived statistics
from raw network traffic flows, lacks the detailed information
crucial for these tasks.
For instance, network analysts often use tools like
Wireshark to investigate network traffic details like packet headers and sequences to diagnose issues or assess performance,
such as tracing latency causes or detecting unauthorized access.
However, the high-level statistical nature of NetFlow data omits
such fine-grained details, rendering it inadequate for such in-depth
analysis.
Furthermore, synthetic NetFlow data cannot be retransmitted
or replayed over network interfaces using tools like tcpreplay.
Retransmitting network traffic is pivotal for various network
testing and validation scenarios, such as evaluating the
performance of network security tools under realistic traffic conditions
or stress-testing network infrastructure.
The absence of packet-level details in NetFlow data precludes
its use for retransmission tasks.
Moreover, certain network analysis tasks require deriving
additional attributes directly from raw network traffic.
For example, estimating the window size or
investigating the distribution of packet sizes across a network
necessitates access to raw traffic data.
These computations are crucial for understanding network
behavior and optimizing network configurations.

Converting high-level statistical NetFlow data back to
raw network traffic, such as packet captures,
is inherently challenging due to the loss of detailed attributes
like header values and flags,
thereby limiting the utility of synthetic NetFlow data for a myriad
of non-ML network tasks. This challenge underscores
the necessity of generating raw network traffic.
At the same time, existing network data generation methods
often neglect to ensure that synthetic data adhere to critical
transport and network layer protocol rules found in real network traffic,
which are crucial for traditional network analysis and testing tools.
For example, a protocol constraint is that packet length frames
must conform to specific sizes as per protocol standards.
Generating synthetic data with incorrect frame sizes could lead
to misinterpretations in data analysis or malfunctions in network
testing scenarios. Other protocol constraints, like correct sequence
numbers in TCP transmissions, valid checksums, and appropriate
flag settings, are also crucial as they affect how network devices
and analysis tools interpret and process the data.
Hence, adhering to protocol rules is essential for the
accuracy and reliability of network analysis and testing tasks,
and serves as a gauge for the quality of synthetic
network data generated.

	\section{Method}\label{sec:method} 

\begin{figure}[tb]
	\centering
	\includegraphics[width=\columnwidth]{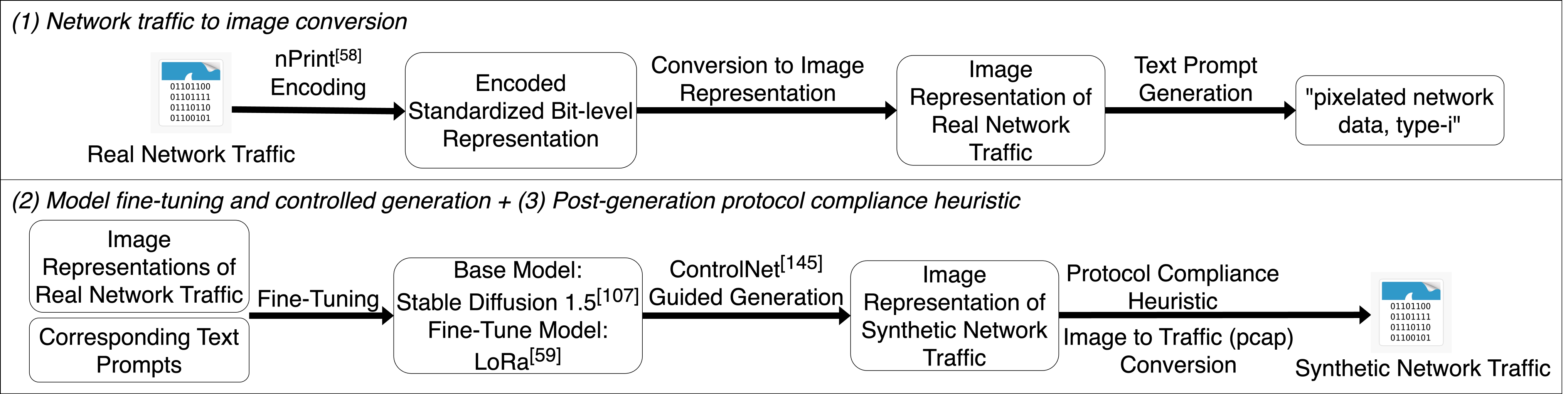}
	
	\caption{Generation Framework Overview.}
	\label{fig:generation_framework}
\end{figure}

In this section, we introduce \emph{NetDiffusion}, a framework that harnesses
controlled text-to-image diffusion models~\cite{rombach2022high} to generate
synthetic raw network traffic that complies with transport and network layer
protocol rules. We find this approach not only elevates classification accuracy
when utilized for data augmentation in ML scenarios but also facilitates a broad
range of network analysis and testing tasks (see Section~\ref{sec:evaluation}),
overcoming the limitations described in
Section~\ref{subsec:inadequate_existing}. We first provide an overview of our
method which has the 3 components shown in
Figure~\ref{fig:generation_framework}, before providing details of each
component.

\noindent \textbf{How do diffusion models work?} Diffusion models synthesize
data by modeling data generation as the process of noise removal from noisy data
(referred to as the reverse process)~\cite{song2019generative,ho2020denoising}.
At training time, a complex ML model, usually a neural network, is trained to
predict noise that is sequentially added to real data (the forward process).
Running the forward and reverse processes in the latent space of a model has
been found to generate better quality data~\cite{rombach2022high}.
Mathematically, consider an initial noise vector (\( \mathbf{z} \)) in the latent
space. The goal of diffusion models is to transform  \( \mathbf{z} \)  into an
data point (\( \mathbf{x} \)) drawn from the desired distribution. The idea is to
define a differential equation that controls the transformation from \(
\mathbf{z} \) to \( \mathbf{x} \) over a series of discrete time steps. The
rationale behind this is that by breaking down the generation process into a
series of incremental diffusion steps, the model can capture intricate
dependencies and details in the data manifold. An essential component of this
approach is score-based generative modeling, where the gradient of the data
log-likelihood with respect to the data (often termed the ``score" function) is
estimated. Modeling the score function is preferable because directly modeling
the probability distribution poses challenges, especially in obtaining the
correct normalization constant~\cite{song2017nice,song2019generative}. Diffusion
models have been adopted most effectively in the context of text-to-image
synthesis, where images matching a given text prompt are to be generated. The
text prompt serves as a conditioning variable to guide the reverse process
towards generating an image that semantically aligns with the text prompt. By
iteratively applying the score function, conditioned on the text prompt, the
model steers the data from a simple prior distribution (like Gaussian noise) to
the desired complex image distribution.
In our case, the text prompts characterize the \textit{specific classes/types of network traffic}
that we aim to generate.

We extend these principles on the operation of diffusion models to network
traffic data via our NetDiffusion framework. Our approach is structured around
three primary components:

\begin{enumerate}
	\item Converting raw network traffic in packet
	capture (pcap) format into image-based representations
	(\cref{subsec:traffic_to_image}).
	\item Fine-tuning a Stable Diffusion model to enable controlled
	 text-to-traffic generation with high distributional similarity across
	 header fields to real-world network traffic 
	  (\cref{subsec:fine_tune_diff}).
	\item Domain knowledge-based post-processing heuristics for detailed
	modification of generated network traffic to ensure high level of protocol
	rule compliance (\cref{subsec:protocol}).
\end{enumerate}

\subsection{Network Traffic to Image Conversion} \label{subsec:traffic_to_image}
In this subsection, we explain the motivation and the process of representing
network traffic as images, an important step in our method.
\subsubsection{Advantages of Using Image Representation of Network Traffic.}
Network traffic data, with intricate inter-packet dependencies and vast
range of attributes, presents a complex landscape that introduces specific
challenges when it comes to accurate representation and efficient learning.
Network traffic data exhibits \textit{high dimensionality},
particularly when using standardized representations such as
nPrint~\cite{nprint}. For instance,
between the IP and TCP headers alone, there is an abundance of fields (\eg, IP
addresses, ports, sequence and acknowledgment numbers, flags, etc.). nPrint
uses a bit-level and standardized representation to have a consistent format for
each packet by accounting for all potential header fields (even if not present
in the original packet). For instance, while a TCP packet won't have UDP header
bits, the nPrint still includes placeholders for these bits. While this ensures
a uniform input structure for ML models, the attribute count per packet often
exceeds a thousand. This high dimensionality introduces computational
bottlenecks for generative models.
	
Additionally, each network traffic trace, considered as a single network
flow/session, inherently contains \textit{sequential dependencies} between
packets. For example, in TCP, packets need to follow a particular sequence to
ensure the integrity and reliability of data transmission. The order of packets,
dictated by sequence and acknowledgment numbers, is crucial to
reconstruct the transmitted data accurately at the receiver's end. These
dependencies are also beneficial to improve ML classification accuracy as they
may be unique to different classes of network traffic. Traditional tabular
formats fall short in preserving these sequential relations due to their static
nature, which could lead to the misrepresentation of the underlying network
behavior~\cite{zhu2021converting,zhong2020image,damri2023towards}.

Since recent strides in synthetic data generation have centered around image
generation~\cite{zhang2022styleswin,chavez2014super,rombach2022high,ramesh2022hierarchical,croitoru2023diffusion},
we seek to leverage these methods to generate high-fidelity synthetic network
data with low computational complexity. Our reasons for adapting these models
are: (1) \textit{Maturity of Image Generative Models}: The advancements in the
domain of image generative models, such as diffusion models, offer a robust
foundation to produce detailed synthetic network traffic. These models have been
optimized over years to understand and reproduce intricate patterns in
high-resolution
images~\cite{rombach2022high,ramesh2022hierarchical,croitoru2023diffusion}. (2)
\textit{Spatial Hierarchies and Connectivity}: Images inherently capture spatial
hierarchies, which is crucial for representing intricate inter-packet and
intra-packet dependencies in network traffic. Pixels in images naturally form
patterns and structures. Deep learning models, especially convolutional neural
networks (CNNs), are adept at exploiting these structures to capture both local
and global dependencies. Unlike traditional tabular formats where data points
might be perceived as independent entities, images inherently emphasize the
significance of a packet concerning its neighboring packets, preserving crucial
contextual
information~\cite{shi2016end,chen2017seq2img,krizhevsky2012imagenet,maas2013rectifier}.
(3) \textit{Visualization and Interpretability}: Image representations offer a
intuitive way to discern packet flows, anomalies, and patterns in network
traffic. (4) \textit{Research and Tools Availability}: The extensive research
and tools available in computer vision mean that scalability and optimization
are already mature, providing a significant advantage when handling
high-dimensional data like network traffic
~\cite{zhu2023conditional,rombach2022high,zhao2023unicontrolnet}.

\subsubsection{Conversion Process.}
To arrive at image representations of network traffic, we first encode packet
captures (pcaps) using nPrint~\cite{nprint}, which converts network traffic into
standardized bits where each bit corresponds to a packet header field bit as
shown in Figure~\ref{fig:generation_framework}. This binary representation is
simple yet effective, where the presence or absence of a bit in the packet
header is denoted as 1 or 0 respectively, and a missing header bit is
represented as -1. This encoding scheme ensures a standardized representation
irrespective of the protocol in use. The payload content is not encoded since it
is often encrypted. However, the size of the packet payloads can be inferred
from other encoded header fields such as the IP Total Length fields.

Following this encoding, a sequence of pcaps is converted into a matrix, which
is then interpeted as an image. The colors green, red, and gray represent a set
bit (1), an unset bit (0), and a vacant bit (-1), respectively. This color
coding provides a visually intuitive representation of the network traffic. Due
to the limitations in the generative models' capability to handle very high
dimensional data, we group packets in groups of 1024. This constraint, while
necessary for the current scope, could be revisited in future work to
accommodate larger groups of packets. Through this process, any network traffic
in pcap format is transformed into an image with dimensions of 1088 pixels in
width and 1024 pixels in height, with each row of pixels representing a packet
in the network traffic flow as shown in Figure~\ref{fig:image_representations}.
Any image in this format can be converted back to pcaps in a straightforward
manner. This representation not only preserves the complexity of the data but
also retains the essential sequential relationships among packets, laying a
robust foundation for the ensuing steps in the NetDiffusion pipeline.

\subsection{Fine-Tuning Diffusion Model and Controlled
Generation}\label{subsec:fine_tune_diff} 

Given a real, labeled network traffic
dataset, we first transform the network traffic flows into their corresponding
image representations as previously described. Leveraging these image-based
representations, we then fine-tune a generative model, specifically a diffusion
model, to produce synthetic network traffic.
\subsubsection{Advantages of Text-to-Image Diffusion Models for Network Traffic Generation.}
The decision to employ diffusion models for image-based network traffic generation
over other generative approaches, such as GANs, is anchored on several compelling advantages:

\begin{enumerate}
	\item \textit{High-Fidelity Generation:}
	Diffusion models excel in capturing and replicating intricate data distributions with
	remarkable fidelity~\cite{ho2022cascaded,sehwag2022generating,pandey2022diffusevae}. This attribute is pivotal, given the complex and
	nuanced patterns inherent in real network traffic. The ability of diffusion models to
	closely mimic these patterns ensures that the synthetic traces
	they produce have high resemblance to real traces.
	
	\item \textit{High-Resolution Image Handling:}
	Diffusion models, through techniques like latent diffusion,
	are adept at generating and managing high-resolution images~\cite{ramesh2022hierarchical,rombach2022high,nichol2021glide}.
	This capability is pivotal for our framework, where the image representation of network traffic
	demands high resolution for accuracy and detail retention.
	While diffusion models can be tailed to handle tabular data directly,
	this may forgo the distinct benefits of image representations,
	such as capturing spatial and sequential intricacies,
	as previously discussed.
	
	\item \textit{Conditional Generation:}
	By allowing for conditional generation based on textual prompts, text-to-image
	diffusion models can be instructed to generate network traffic that mirrors
	specific classes or types, offering an unparalleled fusion of precision and versatility.
	The model learns the relationship between text and image during
	its training phase by adjusting its reverse diffusion trajectory based
	on the given textual prompt~\cite{saharia2022photorealistic,gu2022vector,zhang2023text}. This ensures that the final
	generated image aligns with desired image distribution.
	This becomes invaluable when the need arises to produce specific classes of
	network traffic or to conform to protocol rules and other
	essential network characteristics.
	
	\item \textit{Transparency and Training Stability:}
	The nature of diffusion models ensures a transparent generation process, leading 
	to reproducible results. Such transparency are critical for producing network 
	traces that meet specific patterns or constraints, as it shows good interpretability. 
	Moreover, unlike the often unpredictable training dynamics of GANs due to their 
	adversarial nature~\cite{creswell2018generative,arjovsky2017towards,mescheder2018training}, 
	diffusion models exhibit stable training behavior and well-behaved gradient 
	dynamics~\cite{song2019generative,ho2020denoising,song2020denoising}. This 
	stability not only ensures consistent and anomaly-free output but also streamlines 
	the optimization process.
\end{enumerate}
%

In totality, these advantages make diffusion models a robust and versatile
choice for the generation of synthetic network traces, effectively
addressing the challenges and constraints observed with other generative techniques.

%

\subsubsection{Fine-Tuning a Stable Diffusion Base Model Using LoRa.}
Training generative models, especially those as sophisticated as diffusion models,
from scratch can be resource-intensive and time-consuming.
This is particularly true when considering existing base
models like Stable Diffusion 1.5~\cite{rombach2022high}
which have been pre-trained on the LAION-5B dataset~\cite{schuhmann2022laion}
containing over 5.85B CLIP-filtered image-text pairs.
While the out-of-the-box Stable Diffusion model
is undeniably potent, we cannot directly use it to
synthesize network traffic because
it's designed to cover a broad spectrum of patterns and intricacies,
causing it to lack the depth needed in specific generation tasks.
For instance, generating images based on task-specific prompts
might yield results that, although thematically aligned,
lack the precision and high-fidelity one might expect. 
In our case, a 'Netflix Network Traffic' prompt might
yield a generic image like a highway scene within a Netflix player.
This is particularly evident when the textual description provided has multiple potential visual interpretations, causing the model to produce images that may be blurry or off-target. 
By fine-tuning Stable Diffusion on specific network datasets,
we can address these limitations. Fine-tuning augments the model's
expressiveness, enabling it to better associate with specific patterns
or embeddings of the network traffic domain.

As a result, in this framework, we build upon Stable Diffusion 1.5
and fine-tune this model on our specific network datasets
as shown in Figure~\ref{fig:generation_framework}, making it aptly
suited for generating synthetic network traffic that mirrors the complexities
and nuances of real-world network traffic.
To facilitate this fine-tuning,
we employ Low-Rank Adaptation (LoRa) \cite{hu2021lora},
which is a training technique tailored to swiftly fine-tune diffusion models, particularly in text-to-image diffusion models.
Its crux lies in enabling the diffusion model to learn new concepts
or styles effectively, while maintaining a manageable model file size.
This is beneficial given the traditionally large sizes of models like Stable Diffusion,
which can be cumbersome for storage and deployment.
With LoRa, the resultant models are compact, striking a balance between file size and training
capability. This compactness doesn't sacrifice the model's ability but rather applies minute
yet effective changes to the base/foundational model, ensuring that the core knowledge
remains intact while adapting to new data.

In our fine-tuning process, we start by sampling classes
of real network traffic from our dataset that we aim to
generate synthetically. These traffic samples are then transformed into
their image representations. For each of these images, we craft an unique
encoded text prompt (\eg, ’pixelated network data, type-0’ for Netflix traffic) that succinctly describes
its class type.
The choice of our encoded prompt,
though seemingly simplistic,
achieves two main objectives. It offers a specific vocabulary
that reduces ambiguity and ensures the model hones in
on the network traffic's nuances. Additionally, it minimizes
interference from the base model's original word
embeddings, optimizing the generative process.
Experimentally, we found that this specific prompt structure
provides a balance between specificity and simplicity
to prevent overfitting and misinterpretations, leading to better results.
Subsequently, these image-text pairs are fed into the fine-tuning process,
where the base Stable Diffusion model, augmented with LoRa,
learns to generate network traffic images conditioned on our prompts.
By merging the power of Stable Diffusion models with
the adaptability of LoRa, we create a potent mechanism to generate
high-fidelity, synthetic network traffic images, tailored to our specific requirements.

\subsubsection{Controlled Generation Via ControlNet.}
\begin{figure}[tb]
	\centering
	\includegraphics[width=\columnwidth]{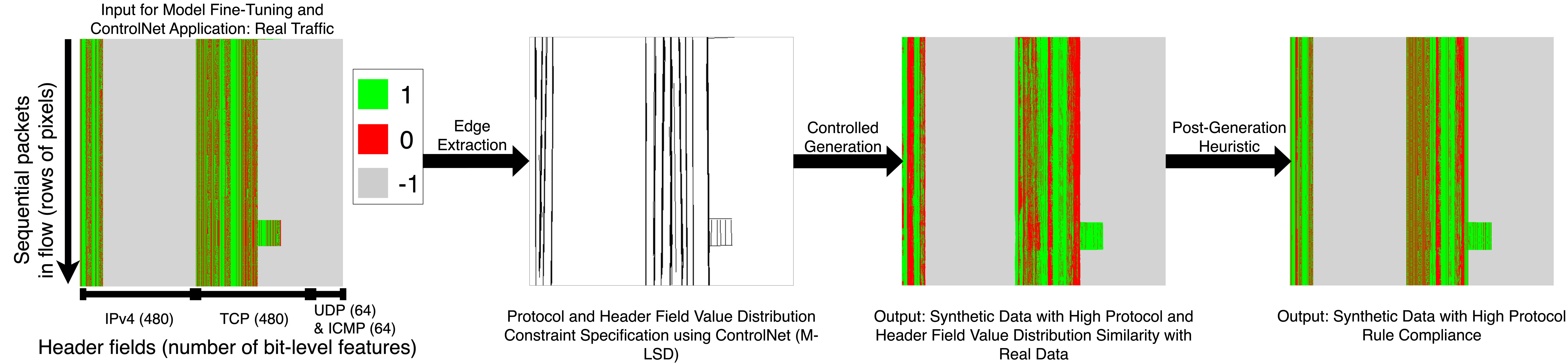}
	
	\caption{Synthetic Amazon network traffic outputs: (1) Using ControlNet,
		we detect regions present in the original traffic and ensure protocol and header field value distribution
		conformance by generating within specified regions. (2) We apply post-generation
		heuristic to refine field details for protocol conformance.}
	\label{fig:image_representations}
\end{figure}

Upon fine-tuning the generation model, the next phase involves generating
the desired class of synthetic network traffic. This is achieved by supplying
the appropriate text prompts to the diffusion models to produce
the image representations of the traffic. Diffusion models operate by simulating
a reverse process from a simple noise distribution to the data distribution,
which enables them to capture and replicate the intricate patterns inherent
in real-world data. The noise is progressively reduced over
several steps, allowing the model to gradually refine the generated image until it
closely resembles genuine network traffic patterns.
We show an example synthetic network traffic in image
representation for Amazon traffic in Figure~\ref{fig:image_representations}.
This prompt-based generation process facilitates the creation
of a synthetic nPrint-encoded network traffic dataset tailored to specific class distribution requirements.
For instance, to curate a dataset with a certain class distribution and size,
one would provide the corresponding quantity of text prompts per class and activate
the generation process accordingly.

However, a challenge arises from the inherent flexibility of general diffusion models.
While they are designed to foster creativity in the generated output,
it can lead to anomalies in the context of network traffic generation.
For example, generated traffic might incorrectly populate packet header fields,
leading to protocol distribution discrepancies between synthetic and real traffic.
Such deviations can compromise ML accuracy and make it arduous to ensure strict
adherence to critical protocol rules as we describe later.
To ensure that the generated traffic closely aligns with the
prevalent protocol and header field value distributions
observed in real traffic, certain constraints are introduced during the generation process.
If, for instance, the actual Amazon network traffic primarily consists of TCP packets,
the generation process should prioritize populating header fields associated with
TCP packets. 
This approach ensures that the generated traffic image predominantly
features green (set bit) or red (unset bit) pixels corresponding to TCP packet headers,
while other pixels remain gray (vacant bit).
Moreover, consistent bit characteristics within headers,
such as consistently unset bits,
should be mirrored in the synthetic output.

Leveraging the controllable nature of diffusion models,
we incorporate ControlNet \cite{zhang2023adding} into the generation process.
ControlNet is a commonly used neural network architecture designed to add spatial conditioning controls to large,
pre-trained text-to-image diffusion models. It capitalizes on the robust encoding layers of these models,
which are pre-trained with vast datasets, to learn a diverse set of conditional controls.
With "zero convolutions", the architecture gradually grows its parameters from an initialized state,
ensuring no adverse noise affects the fine-tuning. ControlNet can work with a range of
conditioning controls, from edges and depth to segmentation and human poses.
It offers flexibility in training, demonstrating robustness with both small and extensive datasets. In our specific use case of ControlNet, we
leverage M-LSD straight line detection for detection the boundaries
between fields that are suppose to be populated and those that are not, as shown in Figure~\ref{fig:image_representations}.
Other edge detection methods such as Canny edge detection
produce similar results. Such line or edge detection methods
are effective because they align with the inherent columnar
consistency present in packet traces.

Incorporating ControlNet allows the synthetic generation process to more closely emulate the
protocol and header field value distributions observed in real network traffic. This minimizes deviations and ensures that
the generated packets largely adhere to the expected protocol type and header field values, enhancing the quality and
reliability of the synthetic network traffic dataset.
And while ControlNet offers coarse-grained control,
determining which image regions to populate,
the diffusion model provides fine-grained control,
specifying individual pixel values which further contributes
to high resemblance to real traffic.

\subsection{Improving Transport and Network Layer Protocol Compliance}\label{subsec:protocol}

Utilizing the controlled diffusion model, we generate encoded network traffic that
adeptly resembles the protocol and header field value distributions
inherent in real-world data.
Our encoded format not only captures every feature observed in real network
traffic but also minimizes the statistical disparity between real and synthetic
feature values, ensuring models can recognize and act on underlying patterns.
Yet, the domain of network dataset augmentation presents unique challenges.
While the synthetic data's quality in ML applications is crucial, its utility extends beyond.
The data's relevance in traditional network analysis and testing tasks
-- often requiring raw network traffic -- becomes equally significant.
Despite the guidance provided by ControlNet during the generation process,
converting our synthetic encoded traffic back to raw formats,
like pcaps, isn't straightforward.
This complexity arises from the multitude of detailed transport  and network layer protocol
rules at both inter and intra-packet levels. Properly formatted traffic must
strictly adhere to these rules. We now explore these complexities and their implications.
\subsubsection{Inter and Intra Packet Transport and Network Layer Protocol Rules.}

At its core, both transport and network layer protocol rules
define the conventions and
constraints that ensure seamless communication between devices in a network.
These rules are crucial as they dictate the structure, formatting, and sequencing of packets,
ensuring that data transmission occurs efficiently and reliably.
While transport layer rules emphasize end-to-end
communication and reliability, network layer protocols focus
on packet routing and address assignment. Combined,
these rules can be broadly divided into two categories:
\begin{enumerate}
	\item \textit{Inter-Packet Rules:} These rules dictate the relationships and sequencing between header fields
	within multiple
	packets in a network flow. For instance, in a typical TCP connection, packets need to be sequenced properly,
	starting with the handshake process involving SYN and SYN-ACK flags. The integrity of data transfer
	is ensured by aligning sequence numbers and acknowledgment numbers.
	Misalignment or incorrect sequencing can disrupt the connection or data transfer process.
	\item \textit{Intra-Packet Rules:} These pertain to the structure and contents within individual packets.
	For example, many protocol headers have a checksum field computed based on the packet's contents to detect
	errors during transmission. It's crucial that the checksum is consistent with the packet's payload.
	Additionally, certain fields within a packet, such as port numbers in TCP and UDP headers,
	must adhere to specific formatting and value constraints to ensure the packet's validity and proper routing.
\end{enumerate}

Ensuring compliance with these rules is vital.
Properly formatted traffic is not only more efficient but also
crucial for network applications devices, such as analysis tools, routers, and firewalls,
which rely on well-structured packets to function correctly.
The challenge with synthetic data generation, especially when optimized for ML accuracy,
is that ML models primarily focus on patterns within features that contribute to classification
or prediction accuracy. These models might overlook intricate protocol rules in favor of
patterns that enhance classification performance. For example, a ML model might
deem certain bit patterns as significant for classifying a particular type of traffic,
even if those patterns violate protocol rules.
While our use of ControlNet aids in approximating the
general protocol and header field value
distributions by ensuring correct field population,
it does not fully capture the nuances of specific bit-level values. 
This disparity between ML optimization and
protocol rule compliance accentuates the necessity for
post-generation adjustments.
Instead of entirely overhauling the ML generation process, which would require embedding a vast amount of rule-bound
constraints derived from domain knowledge, post-generation adjustments offer a more manageable approach to refine the
generated data for protocol compliance. Implementing such detailed control during generation remains a challenging
endeavor, and we envision addressing these complexities in future work.

\begin{figure}[tb]
	\centering
	\begin{subfigure}[tb]{0.49\columnwidth}
		\includegraphics[width=\linewidth]{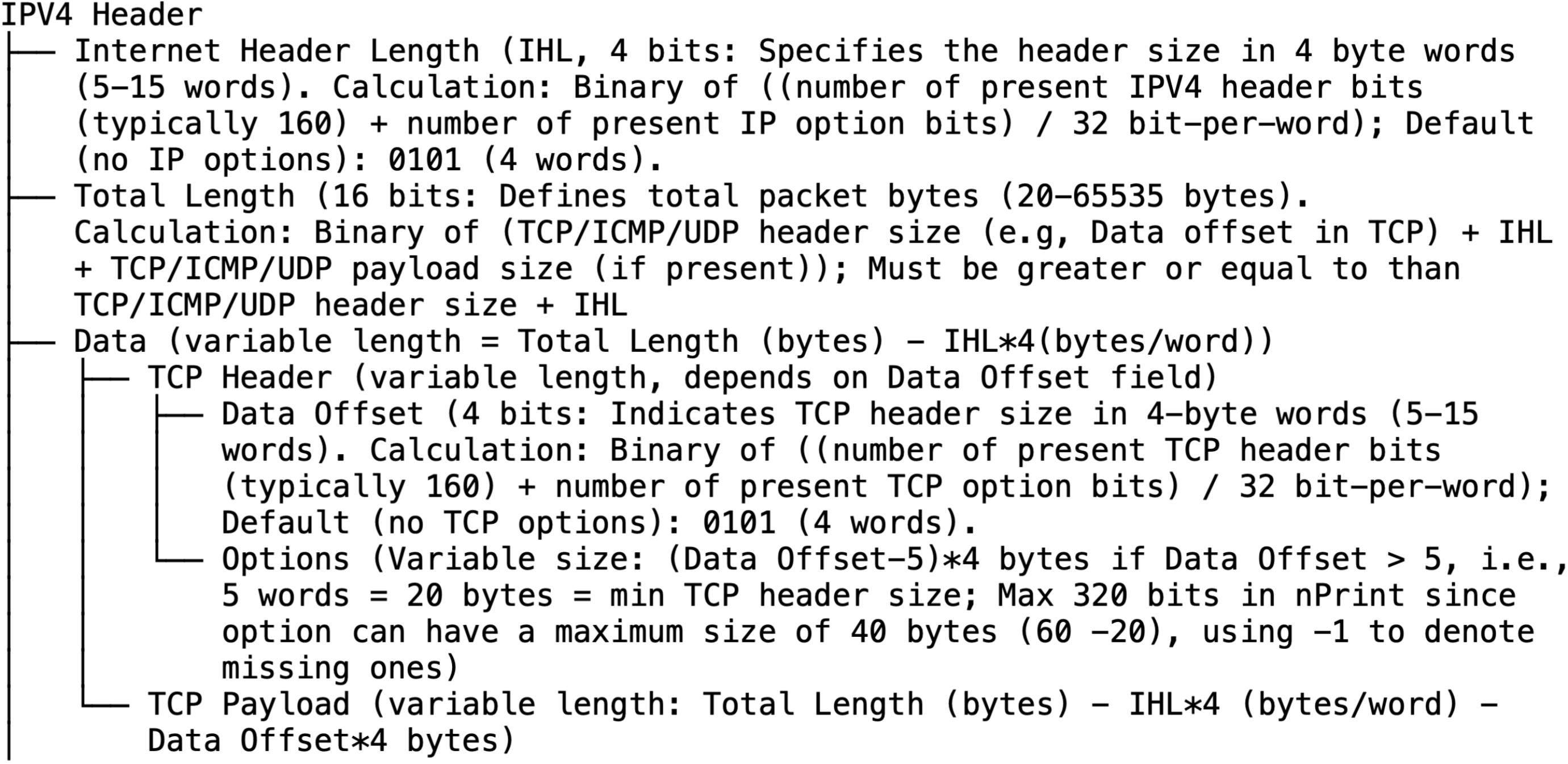}
		\caption{Intra-packet.}
		\label{subfig:tcp_intra_dep}
	\end{subfigure}
	\begin{subfigure}[tb]{0.49\columnwidth}
		\includegraphics[width=\linewidth]{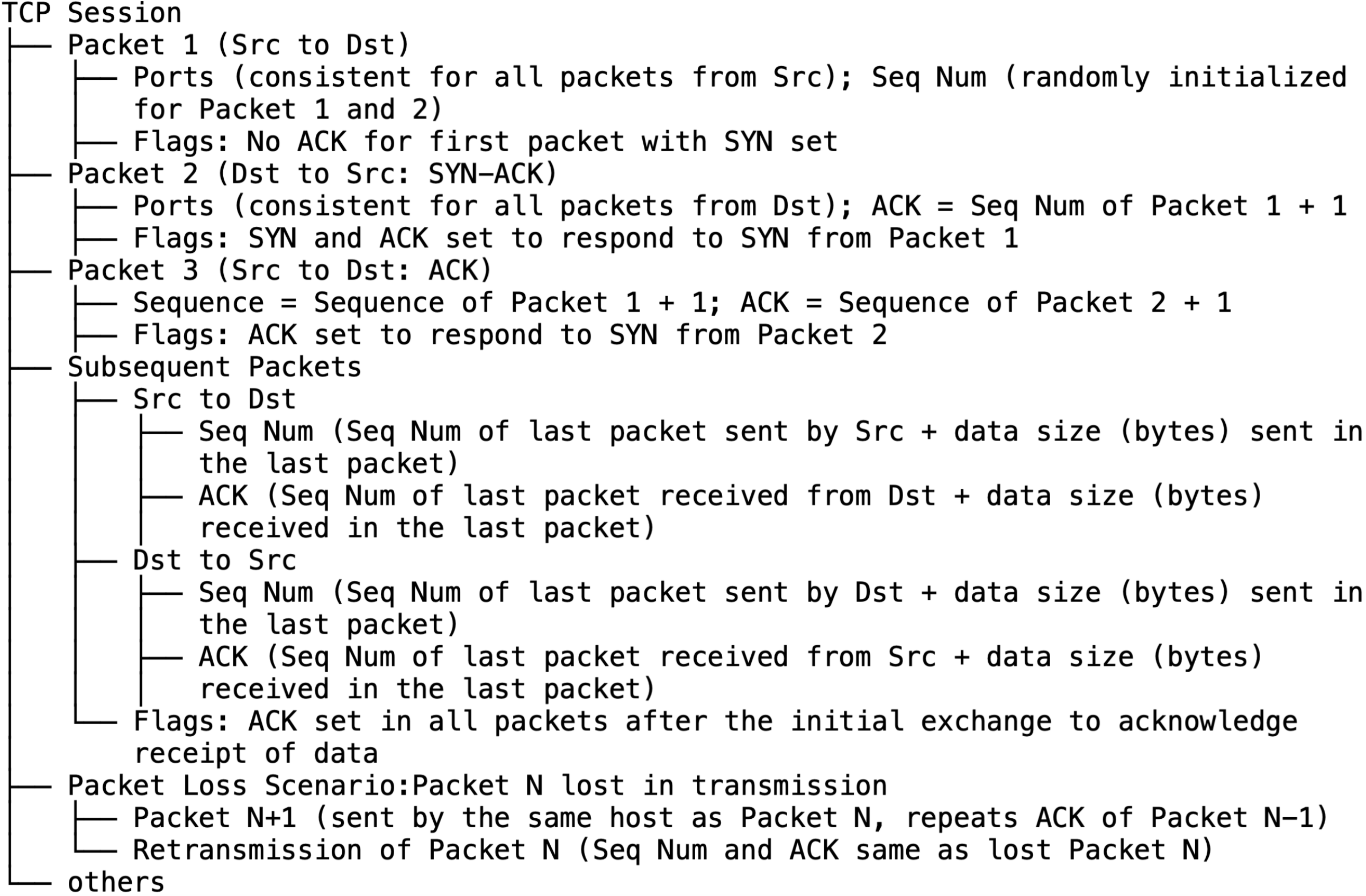}
		\caption{Inter-packet.}
		\label{subfig:tcp_inter_dep}
	\end{subfigure}
	
	\caption{Example TCP protocol rules/dependencies.}
	\label{fig:tcp_dep}
\end{figure}

\subsubsection{Post-Processing Heuristic for Protocol Rule Compliance.}
To maximize the encoded synthetic network traffic's
compliance with transport and network
layer protocol rules,
we first discern a subset of critical header fields that mandate
strict adherence to their formatting rules, \eg,
sequence and acknowledgement numbers. In contrast, some fields can accommodate
a degree of flexibility without compromising the integrity of the network traffic,
such as TCP window size or TTL.
The objective here is to limit the scope of fields requiring modification during
post-processing, ensuring we retain as much of the original generative
model's output as possible.
By doing so, we minimize potential impacts on ML-driven tasks,
while still ensuring the synthetic traffic's compatibility with network analysis tools.
With the critical fields identified, we develop a systematic way to
calculate their correct values based on other generated fields.
This is achieved by constructing two dependency trees—one for intra-packet header field dependencies
and another for inter-packet dependencies.
These trees are built upon domain knowledge and are sourced from standard
network protocol documentation \cite{rfc768,rfc791,rfc792,rfc793,rfc826,beale2006wireshark}.
We present example protocol rules
and the associated dependency trees for
TCP protocol in Figure~\ref{fig:tcp_dep}. More
comprehensive and detailed dependency trees can be
found in the open-sourced repository\footnote{\url{https://anonymous.4open.science/r/packet-capture-dependency-DB0C/README.md}}.

Given a generated encoded network traffic, we begin the correction process
by traversing the trees in an automated, bottom-up fashion. Initially, we satisfy intra-packet dependencies,
ensuring that individual packets are internally consistent.
Subsequently, we address inter-packet dependencies, guaranteeing that the packets
in a flow relate correctly to one another.
Certain fields necessitate uniformity across packets within the same
network traffic trace—like IP addresses and ports.
Others require specific initialization values, such as the IP identification number
and TCP acknowledgment number.
To determine the most appropriate values for these fields,
we employ a majority voting system by selecting the most frequently appearing
value within the generated traffic.
Another notable challenge is timestamp assignment for individual packets,
given its intricate time-series dependencies.
Diffusion models, while adept at spatial dependencies, might struggle with long-range
temporal patterns inherent in time-series.
As a result, our current generation process isn't fully optimized for this.
As an interim solution, we sample original time distributions from real traffic
to produce similar timestamp distribution in the post-generated synthetic data.
We recognize the potential for improvement and plan to address this in future research.
Post these steps, the synthetic traffic should be in a state where it closely
adheres to the essential protocol rules we've identified.
This post-processing ensures that the encoded synthetic traffic can be
seamlessly converted into raw network traffic formats (like pcap) and
subsequently be utilized for a range of non-ML tasks.
	\section{Evaluation}\label{sec:evaluation}
To assess the effectiveness of our generative framework,
we applied it to an exemplary real network traffic dataset,
generating its synthetic counterpart as a case study. Our ML-oriented evaluation comprises
two main analyses: a statistical comparison to gauge the fidelity
of the synthetic data and a model accuracy assessment to determine
its utility in enhancing ML outcomes.
In the non-ML scenarios, we explore the broader implications of
our synthetic data by applying it in diverse network analysis and testing scenarios.
The choice of our baseline dataset, which we detail next, serves as a demonstration
of our method's validity.
\subsection{Dataset Overview and Synthetic Traffic Generation}
\begin{wraptable}[9]{r}{0.5\columnwidth}
	\centering
	\resizebox{0.5\columnwidth}{!}{
		\begin{tabular}{|l|l|l|l|}
			\hline
			\textbf{Macro Services} & \textbf{Total Flows}   & \textbf{Application Labels}& \textbf{Collection Date}  \\
			\hline
			Video Streaming~\cite{bronzino2019inferring}                      & 9465             & Netflix   &   2018-06-01    \\
			&              & YouTube     &       \\
			&              & Amazon    &       \\
			&              & Twitch     &       \\ \hline
			Video Conferencing~\cite{macmillan2021measuring}                        & 6511             & MS Teams         &   2020-05-05    \\
			&              & Google Meet     &       \\
			&              & Zoom     &       \\ \hline
			Social Media~\cite{jiang2023ac}                       & 3610              & Facebook          &   2022-02-08	   \\ 
			&              & Twitter     &       \\
			&              & Instagram     &       \\
			\hline
		\end{tabular}
	}
	\caption{Summary of the real network traffic dataset: 10
		applications across these three macro service types.}
	\label{tab:stats}
\end{wraptable}
Our chosen dataset for real network traffic, summarized in Table~\ref{tab:stats},
comprises pcap files accumulated over different periods, capturing traffic from
ten prominent applications in areas such as video streaming, video conferencing,
and social media.
During preprocessing, we analyze DNS queries to identify relevant IP addresses for
the specified services and applications,
keep only packets associated with these IPs, and
split the traffic into individual flows.
We retain the application and service label for each processed flow,
which is later used for generating text prompts and assessing
classification accuracy\footnote{All traces used in this paper
	are sanitized and contain no personally identifiable information (PII).}.
The comprehensive dataset contains nearly 20,000 flows. For feasibility and consistency
in our evaluations, we randomly sampled 10\% of this collection
We also carried out evaluations on both larger and smaller subsets of the dataset
and obtained comparable results. We adapted the fined-tuned diffusion model to this dataset,
resulting in the generation of a synthetic dataset as outlined in the previous section.
The volume of this synthetic dataset adjusts based on specific evaluation requirements,
as we will detail further. The prompt-driven nature of diffusion model allows for the generation
of synthetic network traffic in any desired quantity, providing flexibility for diverse
analytical needs.
\subsection{Statistical and ML Performance Analysis}
\subsubsection{Statistical Similarity Results.}
A primary measure of synthetic data quality is its statistical
resemblance to the original data. This comparison is critical as
the essence of synthetic data lies in its ability to represent the statistical
properties of the real data without mirroring it exactly. Ensuring statistical
similarity ensures that models trained on synthetic data generalize well to
real-world scenarios.
In our evaluation, we benchmarked our synthetic data against two baselines: the
NetShare method, which produces synthetic NetFlow attributes
and outperforms most of the othe GANs-based methods~\cite{yin2022practical}, and a naive random
generation approach. The latter, by generating purely random values, acts as a
worst-case scenario, illustrating the lower bounds of similarity and underscoring
the value added by more sophisticated methods.
While our diffusion model inherently captures a broader set of network attributes,
for fairness in comparison, we examine similarity both at an aggregated level,
encompassing all features, and at a more focused level, targeting only the common
features between NetDiffusion and NetShare --
using the Protocol attribute as an example.

We employ three distinct metrics to quantify statistical similarity:
Jensen-Shannon Divergence (JSD), Total Variation Distance (TVD), and
Hellinger Distance (HD). JSD gauges informational overlap between distributions,
offering insights into shared patterns. TVD captures the maximum difference between
two distributions, highlighting worst-case discrepancies. Meanwhile, HD, rooted in
Euclidean distance, is especially sensitive to differences in the tails of distributions,
shedding light on distinctions in rare events or outliers. Collectively, these metrics
provide a holistic view of the statistical overlap between the real and synthetic datasets.
Values for all three metrics range between 0 and 1, with values closer to 0 indicating
superior statistical similarity and thus a closer resemblance to the original dataset.


\begin{table}[tb]
	\centering
	\resizebox{\columnwidth}{!}{
		\begin{tabular}{|r|r|r|r|r|r|r|r|}
			\hline
			\multirow{2}{*}{\textbf{Data Format}}&\multirow{2}{*}{\textbf{Generation Method}} & \multicolumn{3}{c|}{\textbf{All Generated Features}} & \multicolumn{3}{c|}{\makecell{\textbf{Ex. Common Feature: Protocol}}} \\
			\cline{3-8}
			&& Avg. JSD & Avg. TVD & Avg. HD & Avg. JSD & Avg. TVD & Avg. HD \\
			\hline
			\multirow{2}{*}{NetFlow} & Random Generation & 0.67 & 0.80 & 0.76 & 0.82 & 0.99 & 0.95 \\
			\cline{2-8}
			 &NetShare & 0.16 & 0.16 & 0.18 & 0.04 & 0.03 & 0.04 \\
			\hline
			\multirow{2}{*}{Network Traffic (pcap)}&Random Generation & 0.82 & 0.99 & 0.95 & 0.83 & 1.00 & 1.00 \\
			\cline{2-8}
			&\textbf{NetDiffusion} & \textbf{0.04} & \textbf{0.04} & \textbf{0.05} & \textbf{0.02} & \textbf{0.03} & \textbf{0.02} \\
			\hline
		\end{tabular}
	}
	\caption{Average normalized statistical differences between real and synthetic network data across (1) all generated fields
		and (2) example commonly generated field -- IPv4 protocol: Jensen-Shannon Divergence (JSD), Total Variation Distance (TVD), and Hellinger Distance (HD).}
	\label{tab:statistical_similarity}
\end{table}

The results in Table~\ref{tab:statistical_similarity} offer a nuanced view into
the challenges and successes of synthetic data generation for network traffic.
At a foundational level, raw network traffic in pcap format is inherently more intricate
than the NetFlow format. This complexity is evident in the stark contrast in statistical
distances when generating synthetic data for these two formats using random methods.
The higher statistical distance observed for the randomly generated raw network traffic
underscores the inherent challenges in replicating its multifaceted nature.
Yet, it's this very complexity that highlights the prowess of NetDiffusion.
Despite pcap being a more challenging format, NetDiffusion—specialized in generating raw
network traffic—demonstrates a remarkable capability. Its statistical distances relative
to real network traffic are notably lower than even NetShare's distances when NetShare is
generating for the simpler NetFlow attributes. This finding underscores the
efficacy of NetDiffusion in producing high-fidelity synthetic data for intricate formats like pcap.

In summary, while the inherent challenges in replicating the complex pcap format
are evident, NetDiffusion's capability to produce synthetic data with high
statistical similarity, even surpassing methods for simpler formats,
validates its potential as a robust tool for data augmentation in the realm of
raw network traffic.

\subsubsection{ML Classification Results.}
To gauge the efficacy of our synthetic network traffic in ML-based data augmentation,
we employ two classification tasks. The first task aims to categorize network flows
at a granular level, aligning them with their corresponding applications (micro-level).
The second task operates at a broader scale, classifying flows into their overarching
services (macro-level). We conduct evaluation using three prominent models,
including random forest (RF), decision tree (DT), and support vector machine (SVM).
We adopt accuracy as the performance metric for these ML-driven
augmentation evaluations due to its intuitive interpretability,
allowing for a straightforward comparison between different augmentation techniques.
Specifically, in scenarios involving classification tasks where classes are
(or are made to be) approximately balanced, accuracy provides a clear picture
of how well the model performs across all classes.
Three distinct augmentation scenarios, utilizing synthetic data, are evaluated:
\begin{itemize}
	\item \textit{Complete Synthetic Data Usage:} Here, either the training or testing set is entirely
	composed of synthetic data, \eg, training exclusively on synthetic data and testing on real
	data, and vice versa. This approach tests the robustness of synthetic data and its capacity to
	emulate real-world data intricacies. Using synthetic data in isolation ensures that models
	are not biased by any inherent patterns of the real dataset during training,
	allowing for an assessment of the synthetic data's standalone quality.
	\item \textit{Mixed Data Proportions:} Synthetic data is interspersed with real
	data at varying proportions, \eg, a 50-50 split between synthetic and real data during
	training. This strategy evaluates the synergy between real and synthetic data.
	Mixing allows models to benefit from the diversity of synthetic data while still
	grounding the learning process in real-world patterns, potentially improving generalization.
	\item \textit{Class Imbalance Rectification:} Synthetic data is employed specifically
	to address and rectify class imbalances in the training set. For instance, underrepresented
	classes in the real dataset are augmented using synthetic data until a balance is achieved.
	This targeted augmentation ensures that the model is exposed to a balanced representation
	of all classes, mitigating biases and improving performance on minority classes.
	Addressing class imbalance is crucial as it prevents models from becoming skewed towards
	overrepresented classes, thereby enhancing their predictive accuracy across all classes.
\end{itemize}

\paragraph{Result on Complete Synthetic Data Usage.}
The results presented in Table~\ref{tab:model_performance}
reveal that our method, which specializes in generating raw network traffic data
in pcap format, consistently outperforms the NetShare approach,
which is tailored for the simpler NetFlow data format.
\begin{table}[tb]
	\centering
	
	\begin{minipage}{0.6\columnwidth}
		\centering
		\small
		\resizebox{\columnwidth}{!}{
			\begin{tabular}{|r|r|r|r|r|}
				\hline
				\multirow{2}{*}{\textbf{Training/Testing}} & \multirow{2}{*}{\makecell[r]{\textbf{Data Format}\\ \textbf{(Generation Method)}}} & \multirow{2}{*}{\makecell[r]{\textbf{Post-Gen.}\\ \textbf{Heuristic}}} & \multicolumn{2}{c|}{\textbf{Highest Accuracy (Model)}} \\
				
				& & &\textbf{Macro-level} & \textbf{Micro-level} \\
				\hline
				\multirow{2}{*}{Real/Real} &Network traffic (N/A) &N/A&1.00 (SVM) & 0.978 (SVM) \\
				\cline{2-5}
				& Netflow (N/A)&N/A&0.86 (RF) & 0.648 (DT) \\
				\hline
				\multirow{3}{*}{Synthetic/Real} &  \multirow{2}{*}{\makecell[r]{\textbf{Network traffic}\\ \textbf{(NetDiffusion)}}}&\textbf{Not Used}&\textbf{0.738 (DT)} & \textbf{0.262 (DT)} \\
				\cline{3-5}
				&&\textbf{Used}&\textbf{0.676(DT)} & \textbf{0.222 (DT)} \\
				\cline{2-5}
				& Netflow (NetShare)&N/A&0.396 (RF) & 0.140 (SVM) \\
				\hline
				\multirow{3}{*}{Real/Synthetic} &  \multirow{2}{*}{\makecell[r]{\textbf{Network traffic}\\ \textbf{(NetDiffusion)}}} &\textbf{Not Used}&\textbf{0.542 (SVM)} & \textbf{0.249 (SVM)} \\
				\cline{3-5}
				& &\textbf{Used}&\textbf{0.529 (SVM)} & \textbf{0.182 (SVM)} \\
				\cline{2-5}
				& Netflow (NetShare) &N/A&0.503(SVM) & 0.102 (RF) \\
				\hline
		\end{tabular}}
		\caption{Across all complete synthetic data usage scenarios, Net-Diffusion augmented dataset yields to higher classification accuracy. 
		Only the top-performing model is displayed.}
		\label{tab:model_performance}
	\end{minipage}
	\hfill
	\begin{minipage}{0.38\columnwidth}
		\centering
		\small
		\resizebox{\columnwidth}{!}{
			\begin{tabular}{|r|r|r|}
				\hline
				\multicolumn{1}{|r|}{\multirow{1}{*}{\textbf{Rank}}}& \multicolumn{1}{r|}{\multirow{1}{*}{\textbf{Real/Real}}} & \multicolumn{1}{r|}{\multirow{1}{*}{\textbf{Synthetic/Real}}}\\\hline
				1  & 3\_tcp\_wsize\_13: 0.0115 & \cellcolor{green!20}1\_tcp\_opt\_92: 0.0125 \\
				
				2  & 3\_tcp\_wsize\_15: 0.0107 & 2\_udp\_cksum\_14: 0.0098\\
				
				3  & 1\_udp\_len\_5: 0.0100 & \cellcolor{green!20}8\_tcp\_opt\_78: 0.0093\\
				
				4 & 1\_tcp\_opt\_24: 0.0099 & \cellcolor{green!20}4\_tcp\_opt\_93: 0.0085 \\
				
				5  & 1\_ipv4\_tl\_5: 0.0090 & 9\_udp\_cksum\_14: 0.0085 \\
				
				6 & 0\_tcp\_opt\_6: 0.0088 & \cellcolor{green!20}5\_tcp\_opt\_93: 0.0084\\
				
				7 & 0\_ipv4\_dfbit\_0: 0.0088 & 5\_tcp\_urp\_8: 0.0082 \\
				
				8 & 1\_tcp\_opt\_6: 0.0086 & 9\_tcp\_cksum\_1: 0.0081 \\
				
				9 & 9\_ipv4\_proto\_3: 0.0085 & \cellcolor{green!20}6\_tcp\_opt\_78: 0.0079 \\
				
				10 & 1\_tcp\_opt\_23: 0.0076 & \cellcolor{green!20}2\_tcp\_opt\_93: 0.0079 \\
				\hline
			\end{tabular}%
		}
		\caption{Macro-level RF feature importance for complete NetDiffusion synthetic data
			usage; Green highlights denote shared header fields
			with the real/real scenario.
			Feature structure: packet\_protocol\_header\_bit.}
		\label{tab:feature_importance}
	\end{minipage}
	
\end{table}

The rich feature space inherent in raw network traffic offers a plethora of
learnable patterns that can bolster model accuracy. With a broader
and more intricate feature set, there's more room for the model to identify and
leverage intricate patterns, nuances, and correlations within the synthetic
network traffic data to enhance
its predictive prowess.
At the same time, the higher statistical similarity between real and our synthetic datasets
(as we observed previously)
implies that the synthetic network traffic data mirrors real-world patterns more closely.
This, in turn, means that features in the real dataset that are pivotal in
distinguishing between network flows are likely to retain their
discriminative power in the synthetic dataset. Such preservation of feature
significance ensures that models trained on synthetic data can generalize
more effectively to real-world scenarios.
Supporting this is our feature importance analysis for the RF
 model in the macro-level classification task
 as shown in Table~\ref{tab:feature_importance}.
 When trained on NetDiffusion synthetic data and tested on real data,
 the RF model exhibited
 a propensity to prioritize features (on the header level) that are also
 critical when both training and testing are done on real data.
 This nuanced focus on specific feature subsets is indicative of the model's ability to
 discern and leverage patterns in the synthetic data that are reflective of real-world traffic.
While we use the RF model for subsequent detailed analysis due to its relatively consistent
performance and interpretability across scenarios,
our later sections will delve into a broader spectrum of model performances.

An additional layer to our analysis pertains to the post-generation heuristic
for enhancing the synthetic network traffic's adherence to protocol rules
while minimally altering the diffusion-generated outputs.
The heuristic affects only $\sim$8\% of the synthetic traffic features
which produces marginal influence on ML performance,
with accuracy reductions ranging from 0.013 to 0.067 across all scenarios.

\paragraph{Result on Mixed Data Proportions.}
We introduce the "mixing rate" to denote the percentage of real data replaced by
synthetic data in the training set. This approach ensures the training set size remains
constant across diverse mixing rates, enabling a clear evaluation of the interplay between
the mixing rate and the resultant model accuracy. Introducing a controlled blend of synthetic
data into real datasets can often enhance model robustness by potentially introducing diverse
patterns, a practice that is considered standard in data augmentation.

\begin{figure}[tb]
	\centering
	\begin{subfigure}[b]{\columnwidth}
		\includegraphics[width=\textwidth]{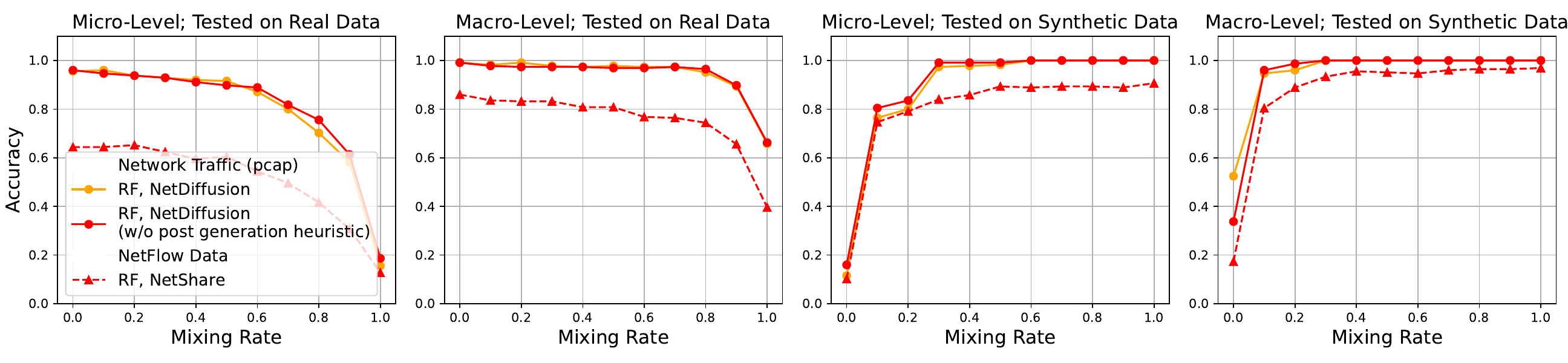}
		\caption{Classification accuracy comparison using the RF model with mixed
			data proportions.
			Datasets augmented with NetDiffusion-generated traffic consistently
			outperform those using NetShare-produced NetFlow attributes.}
		\label{fig:evaluation_compare_simple}
	\end{subfigure}
	
	
	\begin{subfigure}[b]{\columnwidth}
		\includegraphics[width=\textwidth]{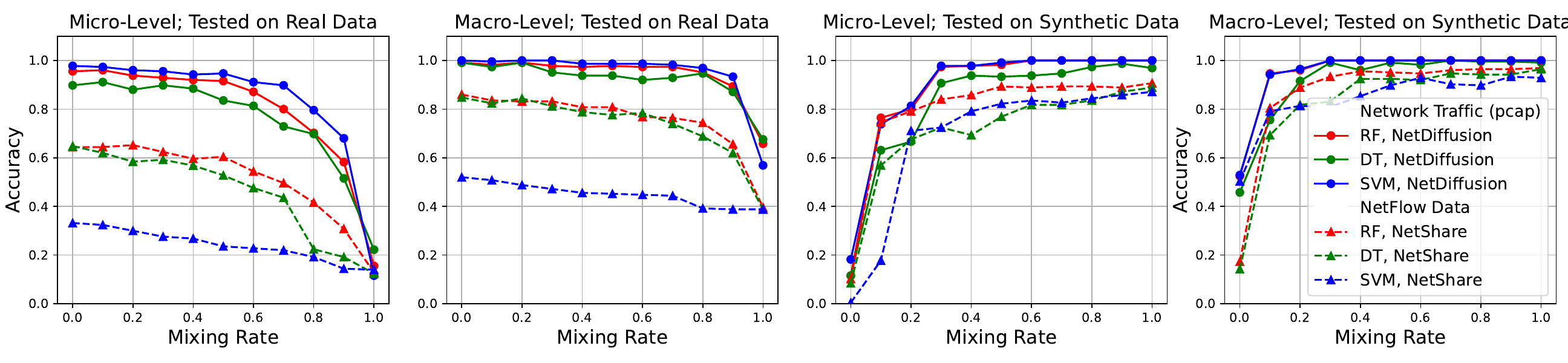}
		\caption{Comparative ML performance across different model choices using
			NetDiffusion-augmented datasets versus NetShare-augmented NetFlow datasets.
			NetDiffusion consistently yields superior results.}
		\label{fig:evaluation_compare}
	\end{subfigure}
	\caption{Evaluation result on mixed data proportions.}
	\label{fig:evaluation_overall}
\end{figure}


Using the RF model as an example, Figure~\ref{fig:evaluation_compare_simple}
shows that models trained with dataset augmented with NetDiffusion-generated traffic consistently achieve higher
classification accuracy than those with NetShare-produced NetFlow attributes.
When testing on real data, models trained entirely on real network traffic demonstrate notably
higher accuracy than those trained solely on real NetFlow (also highlighted in
Table~\ref{tab:model_performance}). This starting accuracy discrepancy is critical. When
integrating synthetic network traffic data into training, any potential degradation in accuracy
is offset by the inherently higher baseline accuracy of real network traffic. In simpler terms,
with a more accurate starting point (real network traffic),
there exists more "buffer" before
accuracy noticeably degrades.
Another pivotal factor is the higher statistical similarity of NetDiffusion-generated traffic
to real data, compared to the similarity of NetShare's NetFlow data to real NetFlow. With this
closer resemblance, as we incorporate more synthetic data into the training, the gradual decline
in accuracy is less pronounced than when using NetFlow data, especially in macro-level classification.
This advantage is not confined to testing on real data. Even when evaluating on synthetic data,
models trained with NetDiffusion's output generally outperform those trained with NetShare.
Additionally,
we carry out a similar analysis as in the case of complete synthetic data usage
by examing the effects of applying our post-generation heuristic on the model accuracy,
as depicted in Figure~\ref{fig:evaluation_compare_simple}.
Consistent with the previous findings from Table~\ref{tab:model_performance},
the example RF model experiences little to no accuracy degredation
as a result of
the post-generation modification.

A notable observation is the sharp decline in accuracy when the data composition of the
training set diverges significantly from the test set. For instance, macro-level classification
accuracy drops when the mixing rate exceeds 0.8. This is expected since adequate samples
from the test data distribution are needed in the training set for effective cross-validation.
As the mixing rate increases, models might overfit to the synthetic data, hindering their
performance on real data. This behavior is evident in the changing feature importance with
increasing mixing rates, as seen in Table~\ref{tab:feature_importance_mixing_rate}.
In practical scenarios, it's rare to rely heavily or solely on synthetic data for training.
Our results suggest that, barring extremes, NetDiffusion-generated traffic can be
effectively used for training.

%
Lastly, we show that across different model choices, as shown in Figure~\ref{fig:evaluation_compare},
NetDiffusion-augmented datasets generally lead to better ML performance than NetShare-augmented
NetFlow datasets.
Notably, the SVM classifier demonstrates markedly superior performance when tasked
with classifying raw network traffic as opposed to NetFlow traffic.
SVMs are intrinsically adept at handling datasets with high dimensionality
and complex relationships between features. The reason lies in SVM's ability
to transform the original data into a higher-dimensional space and find optimal hyperplanes
to segregate different classes. The richer and more intricate the feature space, the more
advantageous this capability becomes.
This observation accentuates the importance of NetDiffusion in generating synthetic
network traffic, which retains the intricacies of real traffic, allowing sophisticated
classifiers like SVM to effectively discern patterns and relationships.

\begin{table}[tb]
	\centering
	\resizebox{0.8\columnwidth}{!}{
		\begin{tabular}{|l|r|r|r|r|r|r|r|}
			\hline
			& Mean (\( \mu \)) & Med & Min & Max & Std Dev (\( \sigma \)) & Range & Var (\( \sigma^2 \)) \\
			\hline
			\textbf{Before Balancing} & $10.00\%$ & $8.89\%$ & $4.44\%$ & $17.78\%$ & $5.13\%$ & $13.34\%$ & $26.37\%$ \\
			\hline
			\textbf{After Balancing} & $10.00\%$ & $10.00\%$ & $10.00\%$ & $10.00\%$ & $0.00\%$ & $0.00\%$ & $0.00\%$ \\
			\hline
	\end{tabular}}
	\caption{Synthetic balancing on under-represented classes to mitigate class imbalance.}
	\label{tab:imbalance}
\end{table}
\paragraph{Result on Class Imbalance Rectification.}
Class imbalance is an ubiquitous challenge in many datasets used for training.
For instance, in our collected network trace, the least represented application class
constituted a mere 4.44\% of the total flow samples,
while the dominant class accounted for 17.78\%,
as shown in Table~\ref{tab:imbalance}. Such imbalances can negatively skew model performance,
as models trained on imbalanced data may struggle to correctly classify underrepresented classes,
especially when real-world test data exhibits a more balanced distribution.
To combat this, a plausible approach is to selectively augment the training set by
appending synthetic data to minority classes,
ensuring an even class distribution.
Simultaneously, by limiting the addition of synthetic data to well-represented classes, 
we minimize the drawbacks associated with integrating synthetic data,
such as the risk of accuracy degradation from overly introduced variations
and insufficient amount of real data in the training set, as we observed
in the case of mixed data proportions.
Using synthetic data offers advantages over traditional methods like SMOTE~\cite{chawla2002smote},
random oversampling~\cite{batista2004study}, ADASYN~\cite{he2008adasyn}, and boosting~\cite{freund1996experiments}. It produces diverse and
novel examples, enriching the feature space and bolstering model
generalization to unseen real-world scenarios. While techniques
like SMOTE replicate close counterparts of real data,
they might miss certain variations.

\begin{table}[tb]
	\centering
	
	\begin{minipage}{0.49\columnwidth}
		\centering
		\small
		\resizebox{\columnwidth}{!}{
			\begin{tabular}{|r|r|r|r|}
				\hline
				\multicolumn{1}{|r|}{\multirow{2}{*}{\textbf{Rank}}}&\multicolumn{3}{c|}{\multirow{1}{*}{\textbf{Mixing Rate (Accuracy)}}}\\\cline{2-4}
				
				& \multicolumn{1}{r|}{\multirow{1}{*}{\textbf{0.0 (0.960)}}}& \multicolumn{1}{r|}{\multirow{1}{*}{\textbf{0.4 (0.928)}}} &  \multicolumn{1}{r|}{\multirow{1}{*}{\textbf{1.0 (0.187)}}} \\\hline
				1 & 5\_ipv4\_ttl\_0: 0.0081 & 1\_ipv4\_ttl\_3: 0.0053 & 5\_ipv4\_ttl\_3: 0.0055 \\
				2 & 4\_tcp\_wsize\_10: 0.0061 & 4\_ipv4\_ttl\_3: 0.0049 &  \cellcolor{red!20}8\_tcp\_cksum\_14: 0.0054 \\
				3 & 5\_tcp\_wsize\_4: 0.0061 & 1\_ipv4\_ttl\_0: 0.0047 &  7\_ipv4\_ttl\_4: 0.0052 \\
				4 & 1\_ipv4\_ttl\_0: 0.0061 & 5\_ipv4\_ttl\_3: 0.0046 &  4\_ipv4\_ttl\_5: 0.0052 \\
				5 & 1\_ipv4\_dfbit\_0: 0.0059 & 4\_ipv4\_ttl\_0: 0.0045  & 3\_ipv4\_tl\_12: 0.0047 \\
				6 & 4\_ipv4\_ttl\_0: 0.0059 &\cellcolor{red!20}7\_ipv4\_tl\_14: 0.0044 & \cellcolor{red!20}1\_udp\_cksum\_4: 0.0044 \\
				7 & 4\_ipv4\_ttl\_3: 0.0055 & 1\_ipv4\_dfbit\_0: 0.0040 & \cellcolor{red!20}6\_ipv4\_tos\_4: 0.0042 \\
				8 & 4\_ipv4\_ttl\_1: 0.0054 & 4\_tcp\_wsize\_10: 0.0039  & 2\_ipv4\_ttl\_3: 0.0041 \\
				9 & 5\_ipv4\_ttl\_1: 0.0053 & 5\_ipv4\_ttl\_1: 0.0038  & 6\_ipv4\_ttl\_3: 0.0040 \\
				10 & 1\_tcp\_opt\_54: 0.0053 & 5\_ipv4\_ttl\_0: 0.0036  & \cellcolor{red!20}8\_tcp\_wsize\_13: 0.0040 \\
				\hline
			\end{tabular}%
		}
		\caption{Feature importance at varying mixing rates for micro-level classification on real network traffic data. Red highlights indicate header fields that are not in the top 10 most important features in the real/real scenario (mixing rate = 0).}
		\label{tab:feature_importance_mixing_rate}
	\end{minipage}
	\hfill
	\begin{minipage}{0.49\columnwidth}
		\centering
		\small
		\resizebox{\columnwidth}{!}{
				\begin{tabular}{|r|r|r|r|}
				\hline
				\textbf{Training Data}&\multirow{2}{*}{\textbf{Model}}&\multicolumn{1}{r|}{\textbf{Test Accuracy}}&\multirow{2}{*}{\textbf{\makecell{$\Delta$ Acc.}}}\\
				\textbf{(Balancing Source)}&  & \textbf{\makecell{Pre/Post Balancing}} &\\
				\hline
				\multirow{6}{*}{\makecell{Network Traffic\\(NetDiffusion)}} & \multirow{2}{*}{\textbf{RF}} &  \multirow{2}{*}{\boldmath\(0.982\rightarrow0.986\)}&\textbf{0.004}\\
				&&&\textbf{Facebook (\boldmath\( 0.955\rightarrow1.00 \))}\\
				\cline{2-4}
				& \multirow{3}{*}{\textbf{DT}} & \multirow{3}{*}{\boldmath\(0.973\rightarrow0.982\)}&\textbf{0.009}\\
				&&&\textbf{Meet (\boldmath\(0.909 \rightarrow 1.00\))}\\
				&&&\textbf{Zoom (\boldmath\(0.955 \rightarrow 1.00\))}\\
				\cline{2-4}
				& SVM & \(0.991\rightarrow0.991\)&0\\\hline
				\multirow{3}{*}{\makecell{Netflow Data\\(NetShare)}} & RF & \(0.645 \rightarrow 0.628\) & -0.017\\
				\cline{2-4}
				& DT & \(0.603  \rightarrow0.600\)& -0.003\\
				\cline{2-4}
				& SVM & \(0.290 \rightarrow0.290\)& 0\\
				\hline
			\end{tabular}
		}
		\caption{Comparison of micro-level classification accuracy: Class balancing using NetDiffusion synthetic data contributes to accuracy improvement on minority classes.}
		\label{tab:imbalance_acc}
	\end{minipage}
	
\end{table}

We apply synthetic balancing using NetDiffusion-generated network traffic,
resulting in a balanced network traffic dataset across all applications.
Similarly, the NetFlow dataset is balanced using synthetic attributes from NetShare.
Our evaluation reveals that models trained on the balanced NetDiffusion dataset
either match or outperform those trained on the original imbalanced dataset
as shown in Table~\ref{tab:imbalance_acc}.
Notably, the accuracy gains were predominantly attributed to the
improved performance on the previously underrepresented classes.
For instance, with the DT model, a notable 0.09 increase in overall classification
accuracy is observed using the NetDiffusion augmented dataset.
A breakdown of this improvement pinpoints Meet and Zoom traffic,
two underrepresented classes with sample counts roughly \textit{half
or less than} the most populated class in the original real dataset,
 as the primary beneficiaries.
Their classification accuracies improved by 0.091 and 0.045, respectively.
In contrast, classifiers trained using the NetShare augmented
NetFlow dataset do not yield such gains and occasionally even
faced accuracy degredations.
This further underscores the higher fidelity of NetDiffusion-generated traffic,
which not only mirrors real data more closely but also supports larger feature
space to enhance model performance.

\subsection{Extendability to Additional Network Anlysis and Testing Tasks}
The efficacy of synthetic data augmentation extends beyond 
just ML performance, especially within the networking realm.
While ML-centric tasks may primarily focus on ensuring that generated, encoded
network traffic produces a consistent feature set with high statistical similarity to real traffic,
conventional network analysis and testing tasks demand the conversion of this generated data
back into raw formats, such as packet captures. Moreover, these tasks require adherence to specific
protocol rules, as elaborated in Section~\ref{subsec:protocol}.
By harnessing the capabilities of ControlNet and the post-generation heuristics we employ,
NetDiffusion facilitates the generation of network traffic that can be seamlessly
converted into raw packet captures while maintaining a robust adherence to protocol rules.
To illustrate this, we use the synthetic Amazon network traffic generated by NetDiffusion
as an example and show that
(1) the generated traffic can be smoothly parsed and interpreted by Wireshark,
a renowned network analysis tool, without encountering exceptions and 
(2)
the synthetic traffic supports retransmission, as corroborated using the
established packet retransmission tool, tcpreplay.
Beyond these observations, we demonstrate
that NetDiffusion-generated traffic can successfully support a broad
spectrum of common network tasks, ranging from intricate traffic analyses to
network behavior studies. Crucially, the derived features routinely employed
in these tasks can be extracted from NetDiffusion's outputs
using Scapy~\cite{rohith2018scapy}.
This underscores the versatility of our approach, suggesting that
NetDiffusion's synthetic network traffic can integrate into a multitude of network
analysis and testing tasks beyond the confines of ML-centric applications.

We abstain from juxtaposing NetDiffusion's capabilities with NetShare in this section.
The rationale is simple: NetShare's design fundamentally restricts it to generating
a limited set of statistical attributes, which inherently curtails its ability to be
converted into raw packet captures, a prerequisite for the tasks discussed here.

\paragraph{Wireshark Parsing Analysis.}
Table \ref{tab:pcap_summary} details the results from Wireshark's parsing of the NetDiffusion
synthetic traffic, stored as capsinfo log~\cite{beale2006wireshark}. Several observations can be drawn:
(1) \textit{Data Format and Integrity:} The generated traffic is stored in the standard pcap
format with Raw IP encapsulation. This confirms the synthetic data's adherence to widely
accepted network trace data formats, ensuring broad compatibility with networking tools.
(2) \textit{Comprehensive Metrics:} All the essential metrics that Wireshark uses to
describe and analyze traffic are present,
ranging from packet count and data size to encapsulation and timing details.
These observations underscore
our design's success in producing protocol rule-compliant synthetic traffic,
ensuring compatibility with analysis tools like Wireshark
that demand structural and semantic correctness.

\paragraph{Tcpreplay's Retransmission Analysis.}
Table \ref{tab:tcpreplay_results} shines light on the retransmission capabilities of the synthetic traffic
via tcpreplay~\cite{tcpreplay2023}. Notable results are:
(1) \textit{Successful Retransmission:} All 1,024 packets were successfully sent without
any failures or truncations, indicating the traffic's high fidelity and adherence to transport layer protocol rules.
(2) \textit{Correct Packet Handling:} Metrics like retried packets standing at
zero and the exact match of unique flow packets to successful packets further reiterate the synthetic
traffic's quality.
(3) \textit{Metrics Completeness:} Key metrics like data bit rate and packet rate, essential for evaluating traffic characteristics, are present and well-defined. 

\begin{table}[tb]
	\centering
	
	\begin{minipage}{0.55\columnwidth}
		\centering
		\small
		\resizebox{\columnwidth}{!}{
		\begin{tabular}{|l|l|}
	\hline
	\textbf{Attribute} & \textbf{Value} \\ \hline
	File type & Wireshark/tcpdump/... - pcap \\ \hline
	File encapsulation & Raw IP \\ \hline
	File timestamp precision & microseconds (6) \\ \hline
	Packet size limit (file hdr) & 65535 bytes \\ \hline
	Number of packets & 1,024 \\ \hline
	File/Data size& 1,335 kB/1,318 kB \\ \hline
	Capture duration & 0.602296 seconds \\ \hline
	First/last packet time (absolute) & 00:00:00.000000/00:00:00.602296 \\ \hline
	Data byte/bit rate & 2,189 kBps/17 Mbps\\ \hline
	Average packet size/rate & 1287.76 bytes/1,700 packets/s\\ \hline
	Strict time order & True \\ \hline
	Number of interfaces in file & 1 \\ \hline
	Interface \#0 info & \begin{tabular}[c]{@{}l@{}}Encapsulation = Raw IP (129 - rawip4)\\ Capture length = 65535\\ Time precision = microseconds (6)\\ Time ticks per second = 1000000\\ Number of stat entries = 0\\ Number of packets = 1024\end{tabular} \\ \hline
\end{tabular}%
}
\caption{Wireshark capinfos validation log on parsing NetDiffusion generated Amazon network traffic.}
\label{tab:pcap_summary}
	\end{minipage}
	\hfill
	\begin{minipage}{0.44\columnwidth}
		\centering
		\small
		\resizebox{\columnwidth}{!}{
	\begin{tabular}{|l|r|}
	\hline
	\textbf{Metric} & \textbf{Value} \\
	\hline
	Total Packets Sent & 1024 \\
	Total Bytes Sent & 1318664 bytes \\
	Duration & 0.613297 seconds \\
	Rate (Bps) & 2150123.0 Bps \\
	Rate (Mbps) & 17.20 Mbps \\
	Packets per Second (pps) & 1669.66 pps \\
	Total Flows & 2 \\
	Flows per Second (fps) & 183.06 fps \\
	Unique Flow Packets & 1024 \\
	Successful Packets & 1024 \\
	Failed Packets & 0 \\
	Truncated Packets & 0 \\
	Retried Packets (ENOBUFS) & 0 \\
	Retried Packets (EAGAIN) & 0 \\
	\hline
\end{tabular}
}
\caption{Tcpreplay results on retransmitting NetDiffusion generated Amazon network traffic.}
\label{tab:tcpreplay_results}
	\end{minipage}
	
\end{table}

\paragraph{Feature Extraction for Core Network Analysis Tasks.}
One of the distinguishing attributes of NetDiffusion generated traffic,
compared to earlier works like NetShare, is the ability to derive detailed metrics
from the synthetic traffic using tools such as Scapy, making it exceptionally valuable
for an expansive range
of network analysis tasks. 
To demonstrate this, we evaluate the applicability of
our synthetic traffic on representative
tasks including traffic and protocol analysis~\cite{kim2006characteristic,barford2002signal,sullivan1998system,zhang2002characteristics,boorstyn1987throughput,dan2006effects},
network performance assessment~\cite{boorstyn1987throughput,dan2006effects},
device identification~\cite{klein2019ip,meidan2017profiliot}, routing and user behavior characterization~\cite{vanaubel2013network,roy2019analysis,gurun2000automating,balachandran2002characterizing,bianco2005web},
and error evaluation~\cite{shannon2001characteristics,jamieson2007ppr,stone2000crc}.
As shown in Table~\ref{tab:traffic_comparison},
using Amazon traffic as an illustrative example,
our synthetic network traffic effectively delivers both fundamental metrics,
such as packet and byte counts, as well as more nuanced measures like the
average TTL used for routing behavior analysis.
Furthermore, metrics linked to network performance, device identification,
routing behavior, and error analysis further accentuate our synthetic traffic's authenticity
and granularity. A noteworthy point is the zero count for error metrics like checksum
errors and fragmented packets in both real and synthetic datasets, underscoring
our synthetic traffic's semantic correctness.
While there are variances between some of the metric values from our
synthetic data and real traffic,
especially in areas like TCP flag distribution, these differences are inherent to
our generative approach.
Instead of merely replicating
real data values, our method generatively produces them, introducing variations
to enhance data diversity.
The delicate balance between maintaining the realism of these variations and the
utility of the resultant metrics for downstream tasks necessitates a nuanced,
task-specific assessment.
In future iterations, we aim to focus on enhancing the congruence 
between synthetic and real data metrics,
such as addressing the overrepresentation of non-ACK packets in the synthetic data
-- a reflection of the complexities tied to emulating the TCP state machine.

In sum, our NetDiffusion generated traffic stands out by allowing the extraction of vital metrics
for additional network tasks,
a capability previous works lacked due to their inability to produce protocol rule compliant,
fine-grained network traffic.
\begin{table}[tb]
	\centering
	\resizebox{\columnwidth}{!}{%
	\begin{tabular}{|r|r|r|r|r|}
		\hline
		Network Task & Metric & Unit & Value (Real Network Traffic) & Value (NetDiffusion Generated Traffic) \\ \hline
		\multirow{3}{*}{\makecell{Traffic Analysis}} & Packet Count & packets & 1024 & 1024 \\
		&Byte Count & bytes & 1100406 & 1318664 \\
		&Avg. TCP Window Size & bytes & 32739.95 & 13234.63 \\\hline
		\multirow{6}{*}{\makecell{Protocol Analysis}} &Protocol Distribution & packets & TCP: 1024, UDP: 0, ICMP: 0 & TCP: 1024, UDP: 0, ICMP: 0 \\
		&\multirow{2}{*}{Flags Distribution} & \multirow{2}{*}{flags} & SYN: 0, ACK: 1023, FIN: 0, & SYN: 68, ACK: 574, FIN: 556, \\
		& & & RST: 0, PSH: 0, URG: 16 & RST: 1, PSH: 6, URG: 495 \\
		&Src Port Distribution & port (packets) & 46508 (303), 443 (721) & 30202 (376), 14443 (648) \\
		&Dest Port Distribution & port (packets) & 443 (303), 46508 (721) & 14443 (376), 30202 (648) \\\hline
		\multirow{2}{*}{\makecell{Network Performance}} &\multirow{2}{*}{Packet Size Distribution} & \multirow{2}{*}{packets} & 0-499: 306, 500-999: 6, 1000-1499: 6& 0-499: 35, 500-999: 56, 1000-1499: 257\\
		&& & 1500-1999: 706, 2000+: 0 & 1500-1999: 676, 2000+: 0 \\\hline
		\multirow{2}{*}{\makecell{Device Identification}} &Src IP Distribution & packets & 192.168.43.37 (303), 54.182.199.148 (721) & 156.76.135.124 (376), 132.81.26.166 (648) \\
		&Dest IP Distribution & packets & 54.182.199.148 (303), 192.168.43.37 (721) & 132.81.26.166 (376), 156.76.135.124 (648) \\\hline
		\multirow{1}{*}{\makecell{Routing Behavior}} &Average TTL & seconds & 186.51 & 123.33 \\\hline
		\multirow{1}{*}{\makecell{User Behavior}} &Number of Sessions & sessions & 1 & 1 \\\hline
		\multirow{3}{*}{\makecell{Error Analysis}} &Checksum Errors & packets & 0 & 0 \\
		&Fragmented Packets & packets & 0 & 0 \\
		&Fragmented IP Datagrams & datagrams & 0 & 0 \\
		\hline
	\end{tabular}}
	\caption{Comparison of Real and Generated Amazon Network Traffic.}
	\label{tab:traffic_comparison}
\end{table}

	\section{Related Work}\label{sec:related}

\paragraph{Diffusion Models}. Since the inception of the first diffusion-based 
model~\cite{sohl2015deep, croitoru2023diffusion}, such models have consistently showcased great 
generative capacities across various domains in image synthesis~\cite{rombach2022high, 
ramesh2022hierarchical}, video creation~\cite{ho2022imagen,ho2022video}, and 
structured data generation~\cite{kotelnikov2022tabddpm,koo2023survey, zhu2023conditional}. 
For instance, Stable Diffusion~\cite{rombach2022high} harnesses pre-trained 
autoencoders for its training, expertly blending detail and simplicity for 
enhanced visual accuracy. DALL-E 2~\cite{ramesh2022hierarchical} employs 
a dual-stage process that crafts realistic images from text-derived embeddings. 
These models often surpass GANs in detail and 
diversity~\cite{karras2020analyzing,kang2023scaling}. While efforts have been 
made to generate structured data such as tabular datasets~\cite{kotelnikov2022tabddpm}, 
producing network traffic presents its distinct challenges due to the strict 
protocol constraints involved. 

\paragraph{Network Traffic Generation}. Traffic generation has been explored 
through various techniques ranging from simulations to GAN-driven machine learning. 
Traditional simulation tools like NS-3~\cite{henderson2008network}, yans~\cite{lacage2006yet},
and the modern DYNAMO~\cite{buhler2022generating} emulate network traffic based 
on different network topologies. Conversely, structure-based solutions~\cite{sommers2004harpoon,
vishwanath2009swing,botta2012tool} capture the network patterns through heuristics, and 
they scale better. A recent trend is GAN-based techniques, exemplified by DoppelGANger\cite{lin2020using} 
and NetShare~\cite{yin2022practical}, which excel in encapsulating intricate temporal 
patterns. However, while traditional methods demand vast domain expertise and 
might lack versatility, GANs, though adaptive, may fall short in protocol 
adherence~\cite{yin2022practical}. Thus these frameworks are not applicable 
to fine-grained traffic analysis tasks like traffic classification~\cite{liu2023amir,jiang2023ac}, 
traffic management~\cite{liu2023leaf,cruz2023augmented}, etc.

\paragraph{Diffusion Process Controls}. Contemporary studies~\cite{zhao2023unicontrolnet} 
aim to harmonize diversity and controllability in data generation by imposing 
conditions. ControlNet~\cite{zhang2023adding}, for instance, integrates task-centric 
conditions into Stable Diffusion, utilizing inputs like edge maps. 
DreamBooth~\cite{ruiz2023dreambooth} employs unique identifiers in text-to-image 
frameworks, ensuring personalized and diverse outputs. Visual ChatGPT~\cite{wu2023visual} 
and DragDiffusion~\cite{shi2023dragdiffusion, pan2023DragGAN}, present an 
interactive image modification system through language or dragging. 
Dall-E 3~\cite{openai2023dalle3} makes a leap forward in to generate images 
that exactly adhere to the semantic meanings of text. 
	\section{Conclusion and Future Work}\label{sec:conclusion}

Synthetic traces, primarily emphasizing certain flow statistics 
or packet attributes, are frequently used to support ML tasks in 
networking. However, their limited alignment with real traces 
and challenges in converting to raw network traffic hinder both 
their ML performance and broader applicability in conventional 
network analyses. 
In our research, we tap into the promising capabilities of diffusion 
models, known for their high-quality data generation, to enhance 
synthetic network traffic production. We present \emph{NetDiffusion}, 
a tool to produce synthetic network traffic, captured as pcaps 
covering all packet headers. Our evaluation reveals that NetDiffusion’s 
pcaps closely resemble authentic data and bolster ML model performance, 
outperforming current methods. These synthetic traces integrate with 
traditional network tools, support retransmission, and suit a broad 
range of network tasks. The rich features in NetDiffusion's outputs 
position it as a vital tool for diverse network analysis and testing 
tasks.
Looking ahead, several avenues beckon further exploration. Our current protocol rule-compliance
approach is post-generative, given the intricate nature of managing inter-dependent
constraints during the diffusion generation process.
A future aspiration is to embed these rules directly within the generation pipeline,
eliminating the need for subsequent adjustments. Additionally, as time dependencies
play a pivotal role, we aim to refine the diffusion models to directly learn and generate
time series, providing a more nuanced approach to inter-packet time dependencies.
Our generation's horizon is presently capped at 1,024 packets per flow sample,
a limitation we seek to address, possibly through techniques like tabular diffusion
that retains packet dependencies or sequential flow generation.
Another intriguing prospect is building a network-specific diffusion foundation model,
which could further heighten generation accuracy.
Lastly, generating semantically meaningful payloads remains a challenge,
with potential solutions like autoencoders offering a promising direction for future work.
	\bibliographystyle{ACM-Reference-Format}
	\bibliography{sigmetrics}

	\appendix
	
\end{document}